\documentclass[pdflatex,sn-mathphy]{sn-jnl}

\usepackage{graphicx}%
\usepackage{multirow}%
\usepackage{amsmath,amssymb,amsfonts}%
\usepackage{amsthm}%
\usepackage{mathrsfs}%
\usepackage[title]{appendix}%
\usepackage{xcolor}%
\usepackage{textcomp}%
\usepackage{manyfoot}%
\usepackage{booktabs}%
\usepackage{algorithm}%
\usepackage{algorithmicx}%
\usepackage{algpseudocode}%
\usepackage{listings}%
\usepackage{rotating}
\usepackage{fancyvrb}
\usepackage{dirtytalk}
\usepackage{comment}
\usepackage{placeins}
\usepackage{enumitem}
\usepackage[numbers]{natbib}
\usepackage{subfigure}
\usepackage{graphicx}
\usepackage{pdflscape}
\usepackage[utf8]{inputenc}
\usepackage{gensymb}

\theoremstyle{thmstyleone}%
\theoremstyle{thmstyletwo}%
\theoremstyle{thmstylethree}%

\begin{document}

\title[Causal Emergence of Consciousness through Learned Multiscale Neural Dynamics in Mice]{Causal Emergence of Consciousness through Learned Multiscale Neural Dynamics in Mice}


\author[1]{Zhipeng Wang}\email{wangzhipeng@mail.bnu.edu.cn}
\author[2]{Yingqi Rong}\email{ryingqi@gmail.com}
\author[1]{Kaiwei Liu}\email{kevinliunxt@mail.bun.edu.cn}
\author[1]{Mingzhe Yang}\email{ymp66393866@163.com}
\author*[1,3]{Jiang Zhang}\email{zhangjiang@bnu.edu.cn}
\author*[4]{Jing He}\email{hejing@shanghaitech.edu.cn}

\affil[1]{ School of Systems Science, Beijing Normal University, Beijing, 100875, China}
\affil[2]{ Johns Hopkins University, 21218, Baltimore, MD, USA}
\affil[3]{ Swarma Research, Beijing, 102308, China}
\affil[4]{ School of Life Science and Technology \& Shanghai Clinical Research and Trial Center, ShanghaiTech University, Shanghai, 201210, China}


\abstract{Consciousness spans macroscopic experience and microscopic neuronal activity, yet linking these scales remains challenging. Prevailing theories, such as Integrated Information Theory, focus on a single scale, overlooking how causal power and its dynamics unfold across scales. Progress is constrained by scarce cross-scale data and difficulties in quantifying multiscale causality and dynamics. Here, we present a machine learning framework that infers multiscale causal variables and their dynamics from near–cellular–resolution calcium imaging in the mouse dorsal cortex. At lower levels, variables primarily aggregate input-driven information, whereas at higher levels they realize causality through metastable or saddle-point dynamics during wakefulness, collapsing into localized, stochastic dynamics under anesthesia. A one-dimensional top-level “conscious variable” captures the majority of causal power, yet variables across other scales also contribute substantially, giving rise to high ``emergent complexity'' in the conscious state. Together, these findings provide a multiscale causal framework that links neural activity to conscious states.}




\maketitle
\section{Introduction}\label{introduction}

Consciousness refers to macroscopic subjective experiences, such as perceiving a scene, enduring pain, entertaining a thought, or reflecting on one’s own awareness~\cite{tononi2016integrated,koch2016neural,tononi2015consciousness}. It disappears during states such as dreamless sleep or general anesthesia, when, from a first-person perspective, all perception fades and experience is entirely absent. A fundamental question that follows is how this macroscopic conscious experience arises from, and relates to, the microscopic activity of individual neurons~\cite{tononi2016integrated}? In recent years, a range of theories~\cite{tononi2016integrated,koch2016neural,seth2022theories,cogitate2025adversarial}—including Global Workspace Theory, Re-entry and Predictive Processing Theories, and Integrated Information Theory (IIT)—have been proposed to address this challenge. These frameworks have laid an important foundation for experimental design, data interpretation, and the development of conceptual and methodological tools to probe the nature of consciousness. Among them, Integrated Information Theory is notable for positing that consciousness corresponds to a maximum of intrinsic cause-effect power, offering a principled framework to quantify both the quality and quantity of experience~\cite{tononi2016integrated,tononi2004information}.

However, the theory ultimately identifies a single complex corresponding to the maximum $\Phi$ that involves only one single-scale (macro) causal power, while not providing a full spectrum of cause-effect strength across scales. This ignores the inherently multiscale causality of the brain, where functional organization extends from individual neurons to microcircuits to large-scale networks~\cite{yuste2015neuron}. Recent advances in causal emergence theory (e.g., CE 1.0 and 2.0)~\cite{hoel2013quantifying,hoel2017map,hoel2025causal} suggest that understanding how causal power measured by Effective Information (EI) varies across different scales is essential for capturing and identifying the phenomenon of emergence. Focusing solely on a single optimal scale is similar to examining a two-dimensional slice of a three-dimensional object; it provides a limited view that risks missing multiscale causal regularities. 

Moreover, IIT overlooks two fundamental and interdependent aspects: first, how information integrates at the neuronal level to generate macroscopic causal power; and second, how this causal power is realized through dynamics. 

Rosas et al. and Luppi et al. showed that the “integration–then–causal realization” process stems from synergistic neuronal interactions, measurable via synergistic information \cite{rosas2020reconciling,luppi2022synergistic,luppi2024synergistic,luppi2024information}. Synergy denotes information that emerges only when multiple regions are considered jointly, underscoring integration’s value. Rosas et al. \cite{rosas2020reconciling} further argued that such neuronal synergy can be abstractly represented by macroscopic, supervenient variables. Adopting this perspective, the present study uses machine learning to derive macroscopic causal variables that integrate information from microscopic neuronal ensembles, while simultaneously capturing the macro-dynamics through which causal power is realized. 
Two distinct modes of information integration and causal transmission may exist: (i) local integration with immediate transmission, which concentrates causal power at the microscopic scale (e.g., cellular automata \cite{wolfram1983statistical}); and (ii) integration via global neural synergy, with transmission through higher-level causal variables, concentrating causal power at the macroscopic scale. This work, therefore, asks: which mode governs the distribution of information integration and causal power?

About the second question, how the multiscale causal power is realized through dynamics. Consciousness is inherently multiscale, yet previous studies have mostly relied on hypothetical or simulated dynamical models~\cite{breakspear2017dynamic,wilson1972excitatory,jansen1995electroencephalogram,friston2003dynamic,siettos2016multiscale}. While insightful, these approaches lack empirical data capturing brain-wide neural dynamics at single-neuron resolution that link micro- and macro-level processes. Traditional recording techniques typically limit research to either high-resolution but spatially localized recordings of small neuronal subsets or low-resolution whole-brain imaging methods like fMRI that cannot resolve neuronal activity~\cite{yuste2015neuron,saxena2019towards,kauvar2020cortical}. Consequently, empirical studies capturing large-scale, high-resolution neural activity across distributed brain regions remain scarce. To address this, we utilize an intravital imaging system that records calcium signals across the entire dorsal cortex of mice at near-cellular resolution, spanning three conscious stages: wakefulness, anesthesia, and post-anesthesia recovery.

This high-quality data enables us to investigate multiscale dynamics and causal effects in a data-driven manner. However, constructing such multiscale dynamical models remains a significant challenge. Several data-driven dynamical modeling methods have been proposed recently \cite{luo2025mapping,d2022quest,gucclu2017modeling,vlachas2022multiscale,scholkopf2021toward,richards2019deep}; however, these approaches either do not account for multiscale causality or do not focus on causal emergence. Consequently, a key gap persists: how to reveal multiscale causality of the dynamics from observational data while rigorously quantifying causal emergence at multiple coarse-grained dimensions. Fortunately, a data-driven multiscale dynamics learning framework, namely Neural Information Squeezer Plus (NIS+), has been proposed to overcome the aforementioned difficulties~\cite{yang2025finding}. It can derive multiscale causal power distributions by maximizing Effective Information for macro-dynamics under the constraint of a high predictive power and has validated the model's effectiveness on both simulated (such as SIR and Boids) and real data (e.g., fMRI data of the brain).

Building on this foundation, we leveraged NIS+ to analyze large-scale, single-cell–resolution cortical imaging recordings from mice across awake, anesthetized, and recovery stages. This allowed us to derive latent causal variables spanning micro-, meso-, and macro-scales, quantify multiscale causal power across different brain states, and reveal underlying causal dynamics and modes of information integration.

\section{Results}\label{results}

\begin{figure}[!htp]
	\centering
    \includegraphics[width=0.95\linewidth]{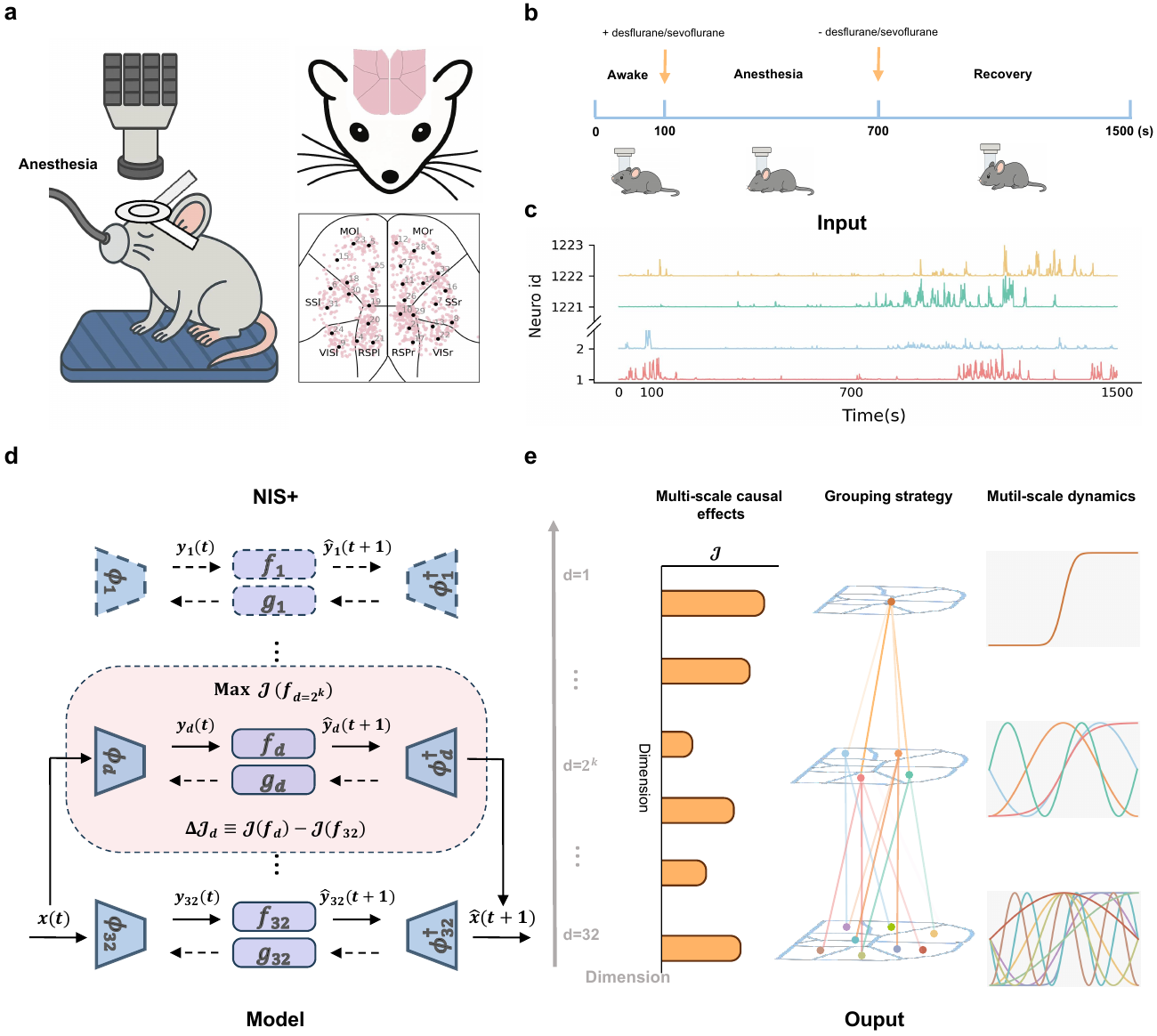}
	\caption{\textbf{A schematic diagram of causal emergence in the mouse brain neural system.} 
    \textbf{a} The schematic diagram of the experimental design, data acquisition setup, and distribution of brain regions. Left panel: Experimental setup showing a mouse head-fixed under a microscope inhaling anesthetic gas (detailed experimental description see Sec.~\ref {experimental_description}). Top right: The schematic diagram of the mouse brain with the regions highlighted in pink. Bottom right: Spatial distribution of eight cortical areas (including: MOl, MOr, SSl, SSr, RSPl, RSPr, VISl, VISr. Abbreviations: MO$-$motor cortex, SS$-$somatosensory cortex, RSP$-$retrosplenial cortex, VIS$-$visual cortex; l and r denote left and right hemispheres). Pink dots represent individual neurons, while black dots indicate neuronal functional ensembles used as input for the NIS+ model's micro-states, obtained by applying K-means clustering to the locations of the neurons (pink dots). \textbf{b} The three stages of anesthesia induction, in order, are: the awake period (0-100s), the anesthetized period (100-700s), and the recovery period (700-1500s). \textbf{c} The schematic diagram of neuronal activity time series (calcium signals). \textbf{d}  Machine Learning Framework: Model Architecture of NIS+. This multiscale framework utilizes the current state $\boldsymbol{x}(t)$ to predict the next state $\hat{\boldsymbol{x}}(t+1)$, which is then compared with $\boldsymbol{x}(t+1)$. While ensuring this prediction error remains sufficiently small, the model learns the dynamics and encoding functions (i.e., the information aggregation strategy illustrated in Fig.~\ref{fig:schematic_diagram}e) at the corresponding scale by maximizing  Effective Information ($\mathcal{J}$) across each dimension. Here, $\boldsymbol{y}_{d}(t)$ and $\hat{\boldsymbol{y}}_{d}(t+1)$ denote the macroscopic causal variable at dimension $d$ at time step $t$ and its predicted value at time step $(t+1)$, respectively. The model comprises four components for each dimension $d$: (1) a forward dynamics learner ($f_{d}$); (2) an inverse dynamics learner ($g_{d}$, used to maximize $\mathcal{J}_d$); (3) an encoder ($\phi_{d}$); and (4) a decoder ($\phi^{\dag}_{d}$). Solid and dashed arrows indicate the forward information flow and the backward path used for training the inverse dynamics, respectively. The pink box denotes the current scale for Effective Information Maximization (detailed model description see Sec.~\ref{model_nis+}). \textbf{e} The output of the model includes: (1) multiscale causal effects ($\mathcal{J}_d$); (2) the information aggregation strategy of the top causal variable (i.e., how $\phi_{1}$ encodes microscopic information); (3) the learned emergent $d$-dimensional dynamics ($f_{d}$).} 
	\label{fig:schematic_diagram}
\end{figure}

To investigate how cross-scale interactions and emergent dynamics shape states of consciousness, we employed Rasgrf2-2A-dCre/Ai148D transgenic mice with sparse GCaMP expression in layer 2/3 pyramidal neurons across the dorsal cortex (Fig.~\ref{fig:schematic_diagram}a). This approach enabled single-cell-resolution calcium imaging across awake, anesthetized, and recovery stages, induced by anesthetics such as desflurane and sevoflurane. Neuronal activity was recorded using a wide-field fluorescence macroscope (6.6 mm field of view), allowing simultaneous monitoring of somatic $Ca^{2+}$ dynamics across visual, somatosensory, motor, and retrosplenial cortices (Fig.~\ref{fig:schematic_diagram}a). During the experiment, mice were head-fixed under the microscope but free to move on a treadmill. After a 30-minute acclimation period, imaging began at time zero with a 100-second awake stage, followed by anesthesia administration from 100 to 700 seconds, and continued through recovery until 1500 seconds (Fig.~\ref{fig:schematic_diagram}b).

Subsequent neuron extraction and data preprocessing yielded multivariate time-series calcium signals (Fig. \ref{fig:schematic_diagram}c). We further aggregated these neurons into several ``neuronal functional ensembles'' (32 clusters as shown in Fig.~\ref{fig:schematic_diagram}a) and fed the aggregated sequence data (details on data generation see Sec.~\ref{data_generation}) into a machine learning framework to reduce the complexity of subsequent analysis and visualization, as well as to lower computational demands. Here, we conceptualize the brain as an autonomous Markovian dynamical system and input the data into the NIS+ framework to learn the brain's dynamics across multiple scales. It is worth noting that increasing the number of clusters did not alter the qualitative results (see  Appendix Fig.~\ref{fig:si_robustness_temporal_coarse_graining}b). Thus, using this 32-dimensional dataset as model input is justified.

Fig.~\ref{fig:schematic_diagram}d demonstrates the architecture of the NIS+ machine learning framework~\cite{yang2025finding}. To enable multiscale analysis, we selected $d=2^k$ for $k=0,1,...,5$ as the dimensions corresponding to different scales, where a higher scale corresponds to a lower dimension. The model architecture at the $k^{th}$ scale consists of four core components: an encoder (${\phi}_d$), a decoder (${\phi}^{\dag}_d$), a forward dynamics learner ($f_d$), and an inverse dynamics learner ($g_d$), all implemented via neural networks. The encoder $\phi_d$ consists of a hierarchy of invertible network layers. Each layer performs a lossless transformation and drops half the dimension~(except $d=5$). The detailed structure of encoders can be referred to Appendix~Fig.~\ref{fig:si_structure_encoders}.

The whole model contains forward and backward information flow paths, which are denoted by solid-line and dashed-line in Fig.~\ref{fig:nis+}, respectively.  Theoretical analysis ensures that by jointly optimizing both paths, our model can predict signals at the next time step while maximizing Effective Information  ($\mathcal{J}_d$, which represents causal power at dimension $d$) at each dimension $d$, thereby enabling effective dynamics learning. 
The forward path consists of information compression via $\phi_d$, causal transformation through dynamics $f_d$, and information generation by $\phi^{\dagger}_d$. We train these networks to minimize the prediction error between $\hat{\boldsymbol{x}}(t+1)=\phi^{\dagger}_d\left(f_d\left[\phi_d\left(\boldsymbol{x}(t)\right)\right]\right)$ and $\boldsymbol{x}(t+1)$. The reverse path trains an inverse dynamics $g_d$ to minimize the error between $\hat{\boldsymbol{y}}_d(t)=g_d\left(\phi_d[\boldsymbol{x}(t+1)]\right)$ and $\boldsymbol{y}_d(t)$. Iterating this bidirectional training yields the causal variables $\boldsymbol{y}_d$, dynamics $f_d$, and the map $\phi_d$ between micro ($\boldsymbol{x}$) and macro ($\boldsymbol{y}$) for $d=2^k, k=0,1,\cdots,5$.

To validate the predictive performance of our multiscale dynamics model, we present Fig.~\ref{fig:dis_of_causal_power}f to show that the Normalized MSE at all dimensions is less than 1, indicating prediction errors smaller than the data variance and confirming the model’s effectiveness. Fig.~\ref{fig:dis_of_causal_power}g compares the training and testing errors before and after data shuffling (data processing is described in Sec.~\ref{data_generation}). The results demonstrate significantly lower errors for non-shuffled data, with only a minimal gap between training and testing errors, suggesting that the model avoids overfitting while capturing genuine and meaningful neural activity patterns. Therefore, subsequent analyses can be reliably conducted based on this trained model.


Upon sufficient training, we derive the $d^{th}$ macro-dynamics $f_d$ (where $d=2^k,k\in\{0,1,\cdots,5\}$) along with its corresponding Effective Information ($\mathcal{J}_d$) and the map $\phi_d$ between micro- and macro-scales (which corresponds to the information integration method) (Fig.~\ref{fig:schematic_diagram}e). See Sec.~\ref{model_nis+} for the detailed method.

Next, we will present the results focusing on four core scientific questions: 1) How is causal power distributed across different dimensions? 2) What are the emergent dynamics learned by the model? 3) How is information integrated across scales? 4) Which brain regions contribute most to the conscious variable?

\subsection{The distribution of multiscale causal power under different stages}\label{causal_emergence}

\begin{figure}[!htp]
	\centering
	\includegraphics[width=1\linewidth]{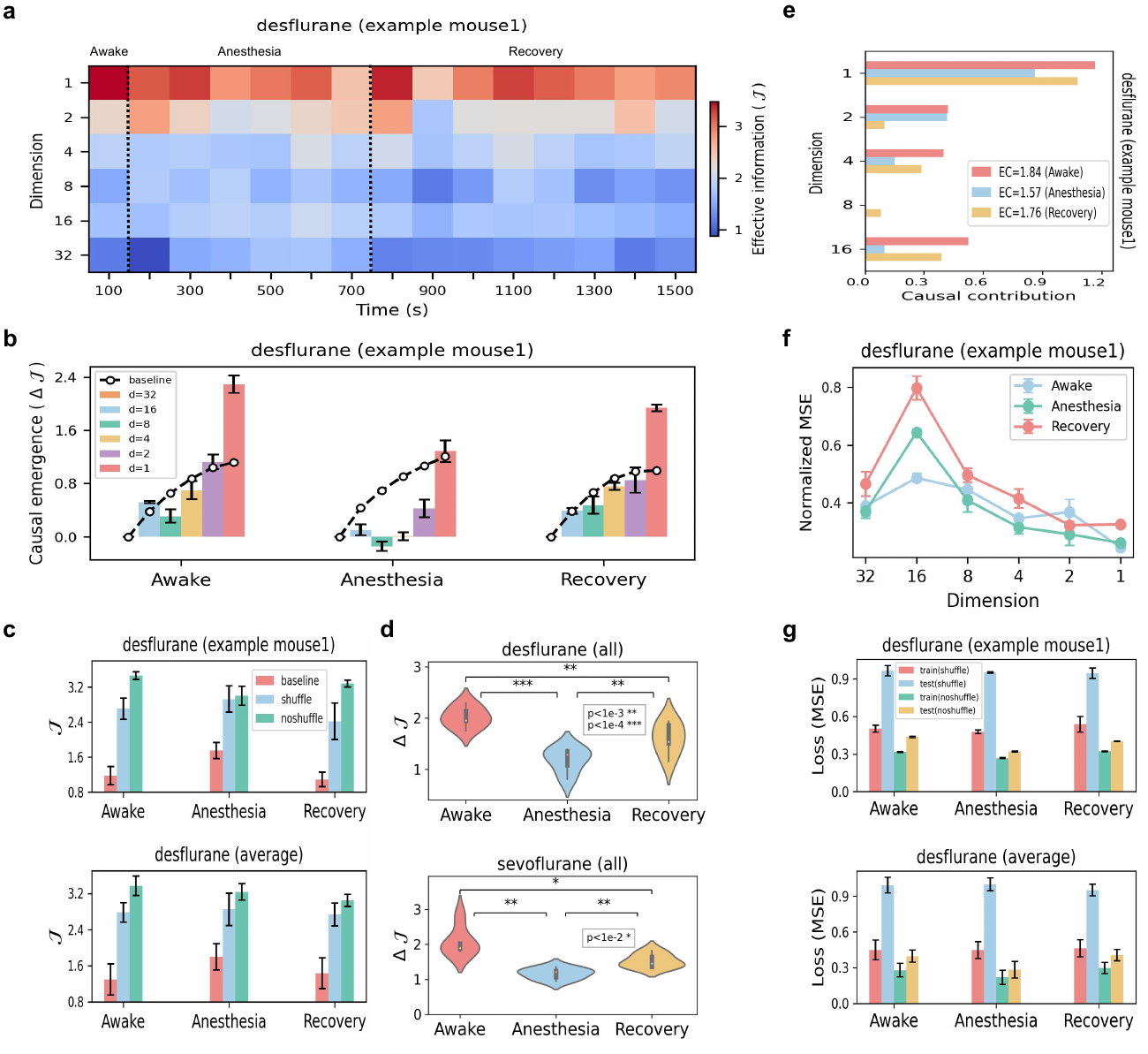}
	\caption{\textbf{The comparison of the distribution of multiscale causal power, degree of causal emergence, causal contribution distribution, and MSE/Normalized MSE across different stages.} \textbf{a} The distribution of causal power ($\mathcal{J}_d$) for mouse1 across different stages and spatial dimensions (where lower dimensions correspond to higher macro-scales). \textbf{b}  The comparison of the degree of causal emergence ($\Delta \mathcal{J}_d$) across dimensions for mouse1 under three different stages. \textbf{c} The comparison of Effective Information ($\mathcal{J}_1$) for mouse1 across three stages at $d=1$ (most macro-scale)— no-shuffled data vs. shuffled data vs. baseline model (Gaussian noise random walk sequence). Here, ``shuffled" indicates random temporal permutation of the data; the ``baseline" model is generated by simulating Brownian motion (with the same variance as the data); and ``average" refers to the mean $\mathcal{J}_1$ across five mice. \textbf{d} The comparison of the degree of causal emergence ($\Delta \mathcal{J}_1 = \mathcal{J}_1 - \mathcal{J}_{32}$) across five mice under three stages and the effects of two anesthetics.  \textbf{e} The comparison of Emergent Complexity (EC) for mouse1 across three stages. EC is calculated based on the causal contribution ($q_k$) distribution, where $q_k = \max(\mathcal{J}_{2^k} - \mathcal{J}_{2^{k+1}}, 0)$ for $k \in \left\{0, 1, 2, 3, 4\right\}$, using Shannon entropy~($H=-\sum_{k}m_k\log_2 {m_k}, m_k= \frac{q_k}{\sum_{k}q_k}$). \textbf{f} The comparison of Normalized MSE (prediction error divided by data variance) on the test set of the NIS+ model for mouse1 across three stages and different dimensions. \textbf{g} The comparison of training and testing errors between the original data and temporally shuffled data across the three stages at dimension $d=1$. The data points and error bars represent the mean and standard deviation of three repeated experiments for the original data and ten repetitions for the shuffled data, respectively.}
	\label{fig:dis_of_causal_power}
\end{figure}

To address the first question, we conducted the following experiments. First, we quantified the causal power in $f_d$ by measuring $\mathcal{J}_d$ across different dimensions (Fig.~\ref{fig:dis_of_causal_power}a). The results show that as dimension $d$ decreases (i.e., the scale becomes higher), $\mathcal{J}_d$ generally increases, indicating stronger causal power at lower dimensions. Fig.~\ref{fig:dis_of_causal_power}c confirms that Effective Information ($\mathcal{J}_1$) at $d=1$ reliably measures the strength of causal effects in the dynamics: the original data showed the highest  $\mathcal{J}_1$  (strongest causal power), temporally shuffled data resulted in reduced $\mathcal{J}_1$ (weaker causal power), and the baseline model simulating a Brownian motion with the same variances as real data, which exhibited the lowest $\mathcal{J}_1$ (see Sec.~\ref{baseline} for baseline definitions).

To quantify the degree of causal emergence on each scale, we calculated $\Delta \mathcal{J}_d\equiv \mathcal{J}_d-\mathcal{J}_{32}$ (Fig.~\ref{fig:dis_of_causal_power}b) for different $d=2^k, k\in\{0,1,\cdots,5\}$ ( $k=5$ is regarded as the micro-level). The study found that across all three stages, the degree of causal emergence ($\Delta \mathcal{J}_d$) peaked at the highest scale (lowest dimension: $d=1$), and the $\Delta \mathcal{J}_1$ of the real data was significantly greater than the Brownian motion baseline, particularly during the awake stage. To further validate the relative magnitudes of $\Delta \mathcal{J}_1$ across the three different stages, we systematically compared all five mice and confirmed statistical significance. As shown in Fig.~\ref{fig:dis_of_causal_power}d, the awake stage exhibited the highest $\Delta \mathcal{J}_1$, while the anesthetized stage showed the lowest. This pattern remained consistent under both desflurane and sevoflurane anesthetics, revealing differences in neural dynamics characteristics across stages. This finding aligns with previous studies: anesthetics suppress high-frequency neural oscillations, enhance low-frequency activity, and reduce neural synchrony, thereby weakening causal interactions~\cite{lewis2012rapid,ching2010thalamocortical}. In contrast, the awake stage promotes cross-frequency coordination, enhances synergistic causal information flow, and strengthens causal effects~\cite{canolty2010functional}.

It is worth noting that, across all stages, the scale with the strongest causal effects is consistently the most macro-scale, which corresponds to the one-dimensional macroscopic causal variable ($\boldsymbol{y}_{d=1}(t)$). This implies that this one-dimensional variable best summarizes the complex dynamics of the mouse brain, where causal effect is maximized without incurring excessive prediction error. We refer to this one-dimensional variable as the “conscious variable”. Integrated Information Theory (IIT) posits that consciousness corresponds to the subsystem that maximizes $\Phi$~\cite{tononi2016integrated,tononi2004information}, which also aligns with the macro-scale where $\mathcal{J}_d$ is maximized—specifically, this one-dimensional macroscopic causal variable. Thus, $\boldsymbol{y}_{d=1}(t)$ can indeed be regarded as a “conscious variable”.

However, even though causal power is maximized at $d = 1$, this does not mean that causal powers at other scales vanish. To more precisely and comprehensively capture the gain in causal power as the scale continues to increase (i.e., as the dimension decreases), we computed the marginal causal contribution at each level (see Fig.~\ref{fig:dis_of_causal_power}e), where the contribution $q_k$ at level $k$ represents the gain in causal power relative to the previous level ($q_k=\max(\mathcal{J}_{2^k}-\mathcal{J}_{2^{k+1}},0), k \in \left\{0,1,2,3,4 \right\}$) and a value of zero indicates no improvement. Furthermore, based on this, we calculated the “Emergent Complexity (EC)”~\cite{hoel2025causal}—defined as the Shannon entropy of the normalized $\left\{q_k\right\}$ (EC=$H(q_k/\sum_{k}q_k$)) as a statistical measure to quantify differences across the three conscious stages. Fig.~\ref{fig:dis_of_causal_power}e shows the causal contribution distribution for mouse1: the awake and recovery periods exhibit a more uniform distribution compared to the anesthetized period, thus demonstrating higher EC. T-tests indicate significant differences between stages (awake vs. anesthetized: $t=4.81, p=0.009$; awake vs. recovery: $t=1.46, p=0.22$; recovery vs. anesthetized: $t=3.42, p=0.027$, where $t$-value reflects the standardized magnitude of inter-group differences, while $p$-value indicates statistical significance). This implies that in the awake stage, causal contributions are more evenly distributed across scales, whereas under anesthesia, they are concentrated mainly at higher scales (lower dimensions), with the recovery stage exhibiting an intermediate pattern.

Therefore, the mouse brain primarily transmits causal power through the ``conscious variable" (the highest scale), while causal contributions are also present across other scales. Secondly, the awake stage exhibits more significant causal emergence and higher EC compared to the anesthetized stage, meaning causal contributions are more evenly distributed across different scales. In contrast, the recovery stage demonstrates an intermediate level of both causal emergence significance and EC, lying between the other two stages.

To verify that our experimental conclusions are independent of the initial choice of the number of neuronal functional ensembles, we doubled the number of the ``neural functional ensembles''. The results show that dimension $d=1$ still exhibits the highest  $\Delta \mathcal{J}_1$, and the ordinal relationship of $\Delta \mathcal{J}_1$ across the three stages remains consistent (see Fig.~\ref{fig:si_robustness_temporal_coarse_graining}b in Appendix~\ref{sec:si_robustness}), strongly confirming the robustness of the conclusions in Fig.~\ref{fig:dis_of_causal_power}. Furthermore, we conducted two additional experiments to validate the model’s robustness: (1) temporal coarse-graining—integrating multiple consecutive time steps as model input; and (2) employing an alternative SVD-based causal emergence identification algorithm~\cite{liu2025svd}. Both experiments consistently observed the highest $\Delta \mathcal{J}_1$ at dimension $d=1$ (refer to Fig.~\ref{fig:si_robustness_temporal_coarse_graining}a and~\ref{fig:si_robustness_temporal_coarse_graining}c in Appendix~\ref{sec:si_robustness}). Additional results for the remaining mice and anesthetics are provided in Appendices~\ref{sec:si_error}–\ref{sec:si_causal_dis}.

\subsection{Analysis of emergent dynamic }\label{dynamic_analysis}

\begin{figure}[!htp]
	\centering
	\includegraphics[width=1\linewidth]{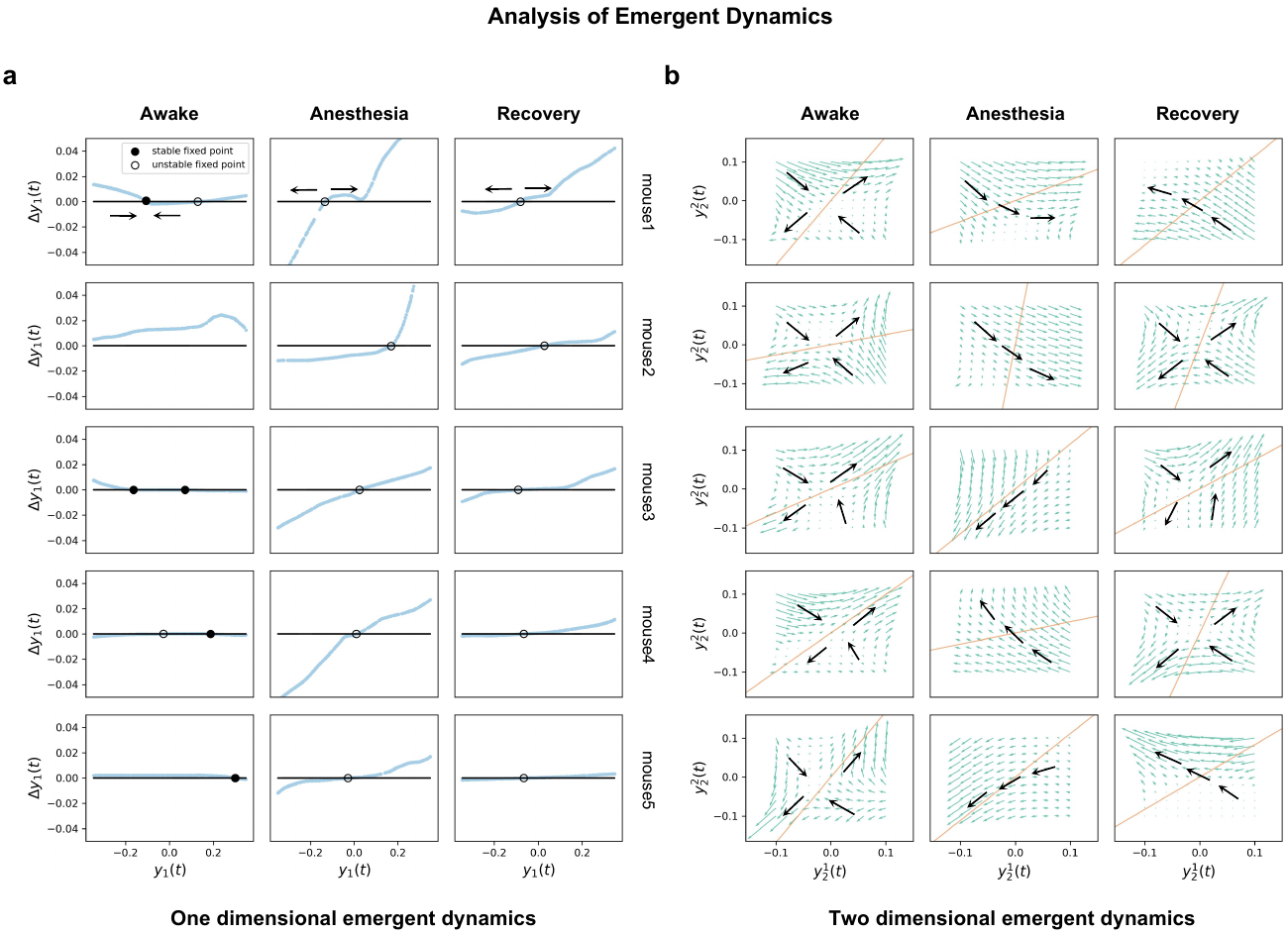}
	\caption{The analysis of macro-dynamics of one-dimensional ($\boldsymbol{y}_1(t)$) and two-dimensional ($\boldsymbol{y}_2(t)$) latent variables for five mice across three different stages, specifically examining the dynamical neural networks $f_d$ for $d=1,2$. The displayed range of the latent variables $\boldsymbol{y}_d(t)$ represents a common subset derived from the projection of all mice onto the corresponding macro-scale. \textbf{a} Phase diagram of the one-dimensional macro-dynamics, illustrating the relationship between the current state $\boldsymbol{y}_1(t)$  and the predicted change $\Delta \boldsymbol{y}_1(t)$ at the next time step. Here, 
     $\Delta \boldsymbol{y}_1(t)$=$\boldsymbol{\hat{y}}_1(t+1)-\boldsymbol{y}_1(t)$,  where $\boldsymbol{\hat{y}}_1(t+1)$=$f_{1}$($\boldsymbol{y}_1(t)$) represents the predicted value of the macroscopic causal variable at time step $(t+1)$. Filled dots denote stable fixed points (points in the neighborhood move toward them), while hollow dots indicate unstable fixed points (points in the neighborhood move away from them). The horizontal black line corresponds to $\Delta \boldsymbol{y}_1(t)=0$. \textbf{b} Vector field representation of the two-dimensional dynamics (see~\cite{yang2025finding} for details). Each green arrow indicates the direction and magnitude of the dynamical derivative ($dy^1_2/dt,dy^2_2/dt$) at the corresponding coordinate points. Black arrows illustrate the overall motion tendency of the vector field. Here shows the case under desflurane anesthesia. The orange line indicates the projection direction that approximately reduces the two-dimensional vector field to the one-dimensional dynamics in Fig.~\ref{fig:dynamics_analyse}a, which is a straight line passing through the origin with a direction determined by $\tilde{\phi}_1$, the last layer of the encoder $\phi_1$ (according to $\boldsymbol{y}_1(t)\approx \tilde{\phi}_1(y^1_2(t),y^2_2(t))$: $\Delta \boldsymbol{y}_1(t) \approx f'_1(\boldsymbol{y}_1(t))=\tilde{\phi'}_1(f_2(\boldsymbol{y}_2(t)))f'_2(\boldsymbol{y}_2(t))$) where $f'_1(\boldsymbol{y}_1(t))$ and $f'_2(\boldsymbol{y}_2(t))$ represent the derivatives of the one-dimensional and two-dimensional dynamics at each state, respectively, and $\tilde{\phi'}_1$ denotes the projection relationship between the two macro-dynamics at each state—specifically referring to the Jacobian matrix of the encoder between the two macro-scales. See Appendix~\ref{sec:si_cal_projection} for further explanation.}
	\label{fig:dynamics_analyse}
\end{figure}

To analyze the learned latent macro-dynamics, we first focus on the dimension with the strongest causal effect ($d=1$), i.e., the dynamical pattern of the ``conscious variable''. We compared the relationship between the one-dimensional macro-variable $\boldsymbol{y}_1(t)$ and its change $\Delta \boldsymbol{y}_1(t)$ (representing the derivative at that point). The results show a nearly consistent pattern across different mice (Fig.~\ref{fig:dynamics_analyse}a). In the awake stage, the dynamics exhibit a stable fixed point, with slow changes around it (manifested as a plateau phase) and a long relaxation time, indicating that the macro-state can stabilize near the fixed point while also switching between multiple states driven by noise or external signals. This aligns with the ability of awake mice to autonomously and flexibly respond to external stimuli and internal intentions~\cite{mcginley2015waking,vyazovskiy2014dynamics,hancock2025metastability}. Regions beyond the plateau, showing upward or downward bending, correspond to unstable brain states. Under anesthesia, the dynamics display only unstable fixed points, and the numerical range is much larger than in the other two stages, suggesting that the conscious variable tends to diverge, exhibiting monotonic increasing or decreasing trends—a hallmark of loss of consciousness. The curve for the recovery stage lies between the awake and anesthetized stages. A plateau phase emerges, indicating very slow changes around the fixed point, yet an unstable fixed point persists. This indicates that the system remains partially destabilized, though significantly less than during the anesthetized stage.

Furthermore, we visualized the learned dynamics of the two-dimensional latent macro-variables ($\boldsymbol{y}_{d=2}(t)$) using a vector field (Fig.~\ref{fig:dynamics_analyse}b). In the awake stage, the vector fields of all mice consistently exhibit saddle points. States near the saddle points may either converge toward attractors or diverge toward instability, which is a characteristic feature of consciousness (\cite{hancock2025metastability,finkelstein2021attractor,tognoli2014metastable,rabinovich2011robust}). Under anesthesia, the dynamics show unstable unidirectional motion, indicating a destabilized state. The recovery stage lies between these two, sometimes displaying saddle-point dynamics and at other times unstable unidirectional motion. These patterns remain consistent across different anesthetics (see Fig.~\ref{fig:si_dyna_analyse_one_two_dim_sev} in Appendix~\ref{sec:si_dynamics_analyse}). Higher-dimensional dynamics do not clearly distinguish among the three stages (see Fig.~\ref{fig:si_dyna_analyse_four_dim_des} and ~\ref{fig:si_dyna_analyse_four_dim_sev} in Appendix~\ref{sec:si_dynamics_analyse}) and fail to reveal unique dynamical mechanisms underlying wakefulness and anesthesia. We further analyzed the relationship between the two macro-scales ($d=1$ and $d=2$), as described by the formula in the caption of Fig.~\ref{fig:dynamics_analyse}b. We find that projecting the two-dimensional vector field along a straight line (the projection direction) can approximate the one-dimensional dynamics shown in  Fig.~\ref{fig:dynamics_analyse}a, and the projection line is derived by assuming a relationship $\boldsymbol{y}_1(t)\approx\tilde{\phi}_1(\boldsymbol{y}_2(t))$ and differentiating it, where $\tilde{\phi}_1$ represents the last layer of the encoder $\phi_1$. In the awake and recovery stages, the projected dynamics often lack a clear unidirectional trend; the saddle points and their neighboring points correspond to the plateau phase observed in Fig.~\ref{fig:dynamics_analyse}a. In contrast, under anesthesia, the projection exhibits a single direction of motion, corresponding to divergent one-dimensional macro-dynamics. Thus, the dynamics of the two panels are generally consistent.

In a word, in the awake stage, the learned one-dimensional ``conscious variable'' exhibits clear metastable dynamics, while the two-dimensional latent variables demonstrate distinct saddle-point dynamics. These findings are consistent with previously reported results in the literature~\cite{mcginley2015waking,hancock2025metastability,tognoli2014metastable}; however, our conclusions are derived from and supported by real experimental data.

\subsection{Information integration}\label{information_integration}

\begin{figure}[!htp]
	\centering
    \includegraphics[width=1\linewidth]{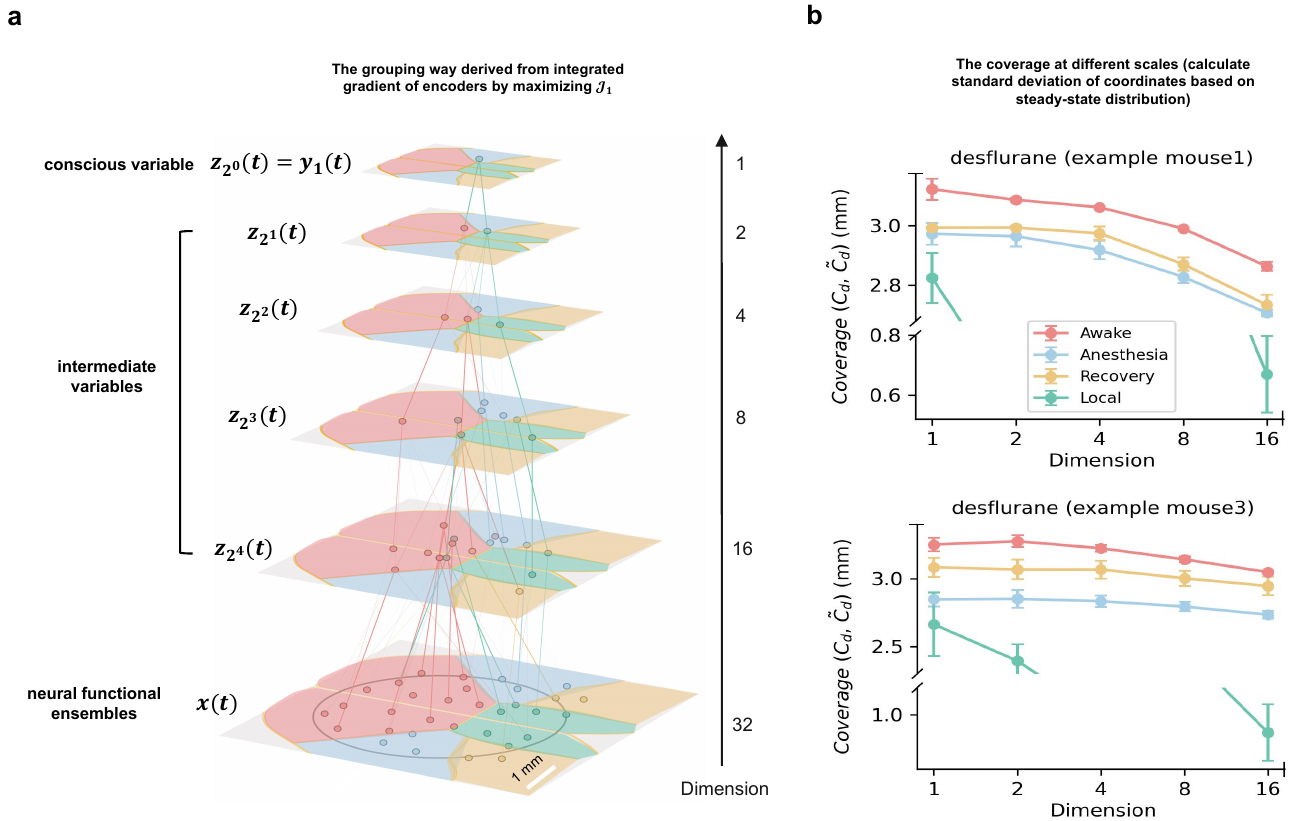}
	\caption{\textbf{The illustration of how the top-scale causal variable learned by NIS+ integrates information from the lower-level variables through attribution analysis, and the coverage of significant attribution areas in the bottom level corresponding to the high-level variables under NIS+ and the hypothetical neighborhood-based aggregation strategy.} \textbf{a} The information aggregation strategy under NIS+. In the panel, hollow nodes represent variables at each layer and dimension, including the neuronal functional ensembles input ($\boldsymbol{x}(t)$), four-layered intermediate variables ($\boldsymbol{z}_{2^{i}}(t)$, $i$ $\in$ $\left\{1,2,3,4\right\}$), and conscious variable ($\boldsymbol{y}_1(t)$). The raw input $\boldsymbol{x}(t)$ is progressively transformed through $\phi_1$ into intermediate variables $\boldsymbol{z}_{2^i}(t)$ of dimension $2^i$ for $i=1$ to $4$, resulting in the final conscious variable $\boldsymbol{y}_1(t)$. A downward edge from a variable indicates the attribution relationship between a node in layer $i$ and a node in layer $(i+1)$. The attribution values are obtained by performing integrated gradient on the $i^{th}$ layer encoder of NIS+ using the values of variables from layer $(i+1)$ as input. The ellipse represents the coverage of significant regions where the conscious variable selected in layer $i=5$ is attributed to the bottom-level neuronal functional ensembles.  The position of each node above 16 dimensions corresponds to the centroid of the cluster formed by the nodes it covers in the layer below, where clustering at each layer is performed using K-means based on spatial coordinates. 
     \textbf{b} The approximate attribution coverages from variables at different dimensions ($d \in \left\{1,..,16\right\}$) down to the bottom layer ($d=32$), obtained by computing the spatial distribution extent of ensembles that significantly  contribution to high-level variables layer by layer (see Sec.~\ref{cal_coverage} for detailed calculation).} 
	\label{fig:information_integration}
\end{figure}

To elucidate how the learned top-scale variable integrates lower-scale information, we conducted attribution analysis on the “conscious variable” $\boldsymbol{y_1}(t)$ to the input variable $ \boldsymbol{x}(t) $ via the hierarchical five-layered encoder $\phi_1$. Starting from the raw input $\boldsymbol{x}(t)$, information gradually passes through $\phi_1$ layer by layer to form a set of intermediate variables $\boldsymbol{z}_{2^i}(t)$ with dimensionality $d=2^i$, $i=1,2,\ldots, 4$, yielding the final variable $\boldsymbol{y}_1(t)$. Here, layer $i = \left\{0,1,\dots,5\right\}$ is numbered from top to bottom. These intermediate variables constitute abstractions of the brain at different scales, as shown in Fig.~\ref{fig:information_integration}a. In which, each small circle denotes one dimension of an intermediate variable. Edges connect arbitrary combinations of dimensions from layer $i$ to layer $(i+1)$. Edge weights indicate attribution strength—the gradient integral~\cite{sundararajan2017axiomatic} of a node in layer $i$ with respect to a node in layer $(i+1)$, averaged over multiple samples.

In this attribution network, we trace $\boldsymbol{y}_1(t)$ top-down through each layer to the input dimensions of $\boldsymbol{x}(t)$. Since $\boldsymbol{x}(t)$ represents neural functional ensembles, this lets us estimate the spatial extent of ensembles that substantially contribute to $\boldsymbol{y}_1(t)$. We model the network as a “Galton board”: propagating $\boldsymbol{y}_1(t)$ downward is like a ball falling through pins, yielding a steady-state distribution of landing positions at the bottom layer. The variance of these positions provides the coverage of $\boldsymbol{y}_1(t)$ over $\boldsymbol{x}(t)$. See Sec.~\ref{cal_coverage} for the detailed method.

Using the same procedure as for $\boldsymbol{y}_1(t)$, we compute the average coverage $C_d$ of the $d$-dimensional causal variable $\boldsymbol{y}_d(t)$ over $\boldsymbol{x}(t)$ and track how $C_d$ changes with $d$. For comparison, we also compute $\tilde{C}_d$ on a hypothetical attribution network built via local information aggregation (see Appendix~\ref{sec:si_hypothetical_grouping_strategy} for details). These results are shown in Fig.~\ref{fig:information_integration}b (top: $C_d$, bottom: $\tilde{C}_d$). We find substantial coverage even at the initial scale ($d=16$), far exceeding the neighborhood-based baseline, confirming distinct integration strategies. Coverage increases as dimension decreases in both networks, indicating that lower-dimensional variables integrate more microscopic information to represent complex macro-states. Coverage is also markedly higher during wakefulness than under anesthesia or recovery, suggesting that conscious processing draws on broader neuronal inputs. Additional results for other mice and anesthetics appear in Appendix~\ref{sec:si_coverage}. Together with Fig.~\ref{fig:dis_of_causal_power}a and Fig.~\ref{fig:information_integration}, these findings indicate that learned lower-leveled causal variables support information integration, while higher-leveled variables exhibit stronger causal properties.

However, it is important to note that the identified integration mechanism does not represent a physically existing neural pathway in the brain, but rather reflects integration patterns learned by the model through latent variables. The macroscopic causal variables can be regarded as a high-level summary of the synergistic interactions among micro-variables. The causal information conveyed by the dynamics of these high-level causal variables also arises from such synergistic interactions, which is approximately quantified as synergistic information in ~\cite{rosas2020reconciling}.

These findings collectively address the fundamental question: ``How does the brain integrate information across scales to transmit macroscopic causal power?"

\subsection{Attribution to Brain Regions}

\begin{figure}[!htp]
	\centering
\includegraphics[width=1\linewidth]{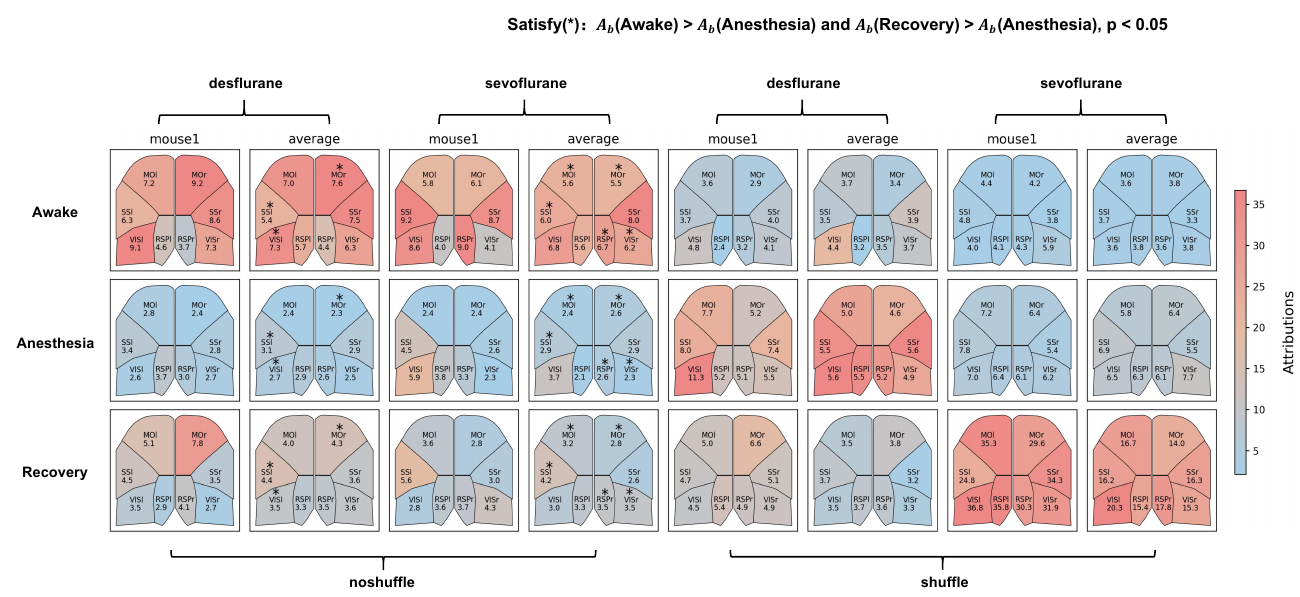}
    \caption{\textbf{The comparison of brain region attribution values for one-dimensional conscious state across three distinct stages (non-shuffled vs. shuffled, desflurane vs. sevoflurane).} The attribution values for each subplot are obtained by summing the attribution values of all neurons within the same brain region, denoted as $A_b$, where $b$ represents the corresponding brain region. Brain regions meeting significance criteria are marked with asterisks ($A_b$(Awake) $\textgreater$ $A_b$(Anesthesia) and $A_b$(Recovery) $\textgreater$ $A_b$(Anesthesia), with $p \textless 0.05$. See text in the upper right corner of the panel.). The attribution values reported in the panel are the mean values of three repeated experiments for the original data and ten repetitions for the shuffled data, amplified by a factor of 1000.} 
	\label{fig:location_brain_area}
\end{figure}

Previous results indicate that consciousness predominantly emerges at the scale exhibiting maximal causal emergence, i.e., at the most macro-scale. To examine which brain regions contribute most to this conscious variable, we compared the attribution results of different brain regions ($A_b$, where $b$ denotes the corresponding brain region) across three stages: awake, anesthetized, and recovery, which is defined as the sum of the attribution values of all neurons within the same brain region and the attribution value of a neuron equals that of its corresponding ensemble (Fig.~\ref{fig:location_brain_area}).

We hypothesized that brain regions contributing critically to the conscious variable would exhibit decreased attribution during loss of consciousness and increased attribution upon recovery, consistently across anesthetics. Fig.~\ref{fig:location_brain_area} presents regional attribution results for a representative mouse and the average across five mice, under both anesthetics and for both non-shuffled and shuffled data. In the non-shuffled data, most brain regions show higher attribution values during the awake stage (red), lower values under anesthesia (blue), and intermediate values during recovery. This pattern is preserved across different anesthetics. Notably, many regions exhibit a significant decrease in attribution from awake to anesthetized stages and a subsequent increase from anesthetized to recovery stages. Regions satisfying these criteria (*$A_b$(Awake) $\textgreater$ $A_b$(Anesthetized) and $A_b$(Recovery) $\textgreater$ $A_b$(Anesthetized), $p<0.05$) are marked with an asterisk and are observed exclusively in the non-shuffled data. Examples include MOr, SSl, and VIS, which show consistent patterns across both anesthetics and may play key roles in shaping the macro-scale conscious variable.

Overall, this attribution analysis, by linking microscopic neuronal ensembles to the macro-scale conscious variable, reveals distinct brain-wide contribution patterns across conscious states (most regions follow $A_b$(Awake) $\textgreater$ $A_b$(Recovery) $\textgreater$ $A_b$(Anesthetized)) and identifies potential consciousness-related regions (e.g., MOr, SSl, and VIS). Attribution results for the remaining mice are provided in Appendix~\ref{sec:si_attri_analyse}.

\section{Discussion} \label{discussion}

Understanding how conscious experience emerges from collective neural dynamics remains a formidable challenge in neuroscience. By analyzing the relationship between multiscale causal power distribution and the emergence of consciousness in mice across awake, anesthetized, and recovery stages, this study yields the following key findings: 1) Causal powers (measured by EI) concentrate in the highest-scale ``conscious variable''. At the same time, causal variables at other scales also exhibit certain contributions, which are most diverse in the awake stage. Thus, two metrics, causal emergence ($\Delta \mathcal{J}$) and emergent complexity (EC), distinguish the three stages. 2) The ``conscious variable'' exhibits metastable dynamics during the awake stage, whereas under anesthesia, the dynamics collapse into an unstable pattern. The recovery stage displays an intermediate behavior. Moreover, the metastable dynamics of this ``conscious variable'' are consistent with the saddle-point dynamics observed in two-dimensional causal variables. 3) Attribution analysis reveals a scale-dependent functional dissociation: micro-scale variables primarily support information integration, while macroscopic causal variables are mainly responsible for transmitting causal information.

Although this study derives a set of multiscale causal variables through machine-learning methods, these discovered variables can also be compared with the findings regarding synergistic and redundant information by Rosas~\cite{rosas2020reconciling} and Luppi et al.~\cite {luppi2022synergistic,luppi2024synergistic}. First, neurons tend to form synergistic information via higher-order interactions to transmit greater causal power at macro-scales, corresponding to our high-scale causal variables. Second, during wakefulness, causal power is also transmitted via local neuronal interactions across multiple scales. These correspond to the model's low-scale causal variables, which arise from local low-order interactions that likely help mitigate noise and uncertainty, thereby enhancing overall dynamical stability. Due to the presence of substantial underlying noise and repetitive information, these low-scale variables may primarily transmit redundant information (refer to Luppi et al.~\cite{luppi2022synergistic}). In summary, the awake brain operates as a complex system in a metastable state: it not only transmits causal information at high scales through extensive synergistic interactions but also increases redundancy via local low-order interactions to improve systemic stability.

Additionally, our theoretical framework offers the following potential contributions to consciousness research:

This study, by comparing the magnitude of causal power across different stages and scales, provides the first empirical evidence revealing the relationship between consciousness and causal emergence. The attenuation of macroscopic causal effects under anesthesia indicates that conscious states depend on the activity level of macroscopic causal variables and the strength of causal power. In other words, the emergence of consciousness requires not only the occurrence of causal emergence in brain dynamics but also that its significance exceeds a certain threshold. Thus, if macroscopic causal power is insufficient relative to the micro-level, the system struggles to sustain unified and coherent cognitive experiences. Notably, $\Delta \mathcal{J}_1 > 0$ was observed across wakefulness, anesthesia, and recovery, indicating that causal emergence per se does not necessarily correspond to a fully unified conscious experience; the positive values during anesthesia may merely reflect residual or low-level awareness. In the awake state, one-dimensional macroscopic metastable dynamics capture the coordinated integration of macroscopic causal variables, which underlie highly unified conscious experiences, whereas this pattern is suppressed under anesthesia. At a broader scale, high-scale causal emergence, together with a diverse distribution of causal contributions across scales, underlies the duality of consciousness—encompassing both flexibility and stability.

Moreover, the large-scale, single-cell resolution cortical imaging data of mice utilized in this study provides a significant contribution to the research on multiscale dynamics of the brain. Meanwhile, the machine learning framework (NIS+) employed in this study shows broad application potential. By automatically learning multiscale dynamics and information integration strategies from observational data, this method overcomes previous methodological limitations. Since such macro-variables and their dynamics are difficult to observe directly in the real brain, this data-driven approach constructs a multi-level surrogate model of the brain that learns causal variables and dynamics across scales. By analyzing within- and cross-level information transfer, it reveals the brain’s information integration strategies and dynamic properties—transforming previously unmeasurable information into quantifiable forms. The model can also train multi-level dynamics directly from neuronal-level micro-states and be extended to neural data at arbitrary spatiotemporal resolutions, such as EEG and fMRI. For instance, it could be used to segmentally model sleep dynamics from EEG signals, enabling sleep-depth quantification and quality monitoring.

However, it is important to emphasize that the causal variables derived through machine learning do not correspond to physically existing entities in the brain. This raises an intriguing question for future research: Could the brain inherently employ information aggregation mechanisms similar to the NIS+ architecture, or even contain neural units capable of implementing macroscopic causal power? It is worth noting that if the brain does utilize analogous mechanisms for information integration and causal transmission, it could form a self-reference system—using local neural circuits to generate a summarized representation of its own global state. Such self-referentiality~\cite{laukkonen2025beautiful} may serve as a fundamental basis for the emergence of consciousness.

Although this study offers valuable insights into causal emergence in neural dynamics, several limitations remain:
(1) Model training requires a sufficiently large amount of low-noise data, and inadequate training may lead to inaccuracies in learned dynamics, thereby affecting subsequent analyses; (2) The model relies on the Markov assumption of dynamics, which may oversimplify neural processes involving memory effects; (3) Since the model design is not inspired by brain neural mechanisms, its internal representations and decision processes are difficult to align with human cognition, limiting functional interpretability; (4) Causal structures between brain regions have not been characterized. Addressing these limitations and exploring potential improvements will help extend the applicability of our theory to a broader range of neural data. These issues await further investigation and resolution in future research.

\bibliographystyle{unsrt}
\bibliography{biblio}

\clearpage

\section*{Methods} \label{method}

\subsection*{Experimental Description}
\label{experimental_description}

\textbf{Animal and surgeries}. All animal procedures were approved by the Administrative Panel on Laboratory Animal Care at ShanghaiTech University. Mice were housed under standard conditions: temperature at 24$\degree$C, relative humidity at 50\%, and a reversed light/dark cycle with lights on from 7:00 am to 7:00 pm. Five male mice used in this study were generated by crossing Rasgrf2-2A-dCre mice (JAX 022864) with Ai148(TIT2L-GC6f-ICL-tTA2)-D mice (JAX 030328), resulting in Cre-dependent expression of GCaMP6f (Jackson Laboratory). Animals were group-housed in plastic cages with standard bedding under controlled room conditions. Experiments were conducted on healthy adult male mice (8–12 weeks old, 20–30 g). To induce Cre activity, trimethoprim (TMP; 0.25 mg/g, Sigma) was administered via intraperitoneal injection for two consecutive days. Mice had ad libitum access to food and water prior to surgery. Following surgery, animals were individually housed for experimental procedures. After recovery from craniotomy, mice were head-fixed on a commercially available passive treadmill equipped with an encoder (Labmaker) for imaging sessions. All animals underwent habituation to the head-fixed setup for at least one week prior to data acquisition.

\textbf{Craniotomy operation for wide-field imaging}. Prior to wide-field imaging surgery, the position of the bregma was recorded using a fixed probe. The probe was then removed to provide sufficient space for cranial removal. A section of the skull was carefully removed and replaced with a sterile cranial window, which was secured using Krazy glue (Elmer’s Products Inc). Wide-field imaging was employed to record single-neuron activity across the entire cortical surface. Next, a head-post was cemented onto the edge of the skull and reinforced with a layer of dental acrylic. Postoperative analgesia and anti-inflammatory treatment were administered by subcutaneous injection of flunixin meglumine (1.25 mg/kg; Sichuan Dingjian Animal Medicine Co.) once daily for five consecutive days. Following craniotomy, mice were allowed a recovery period of at least seven days. After recovery, the integrity of the cranial window and the quality of calcium signals were evaluated. Only mice that fully recovered and exhibited good health were included in subsequent experiments. Notably, mice have been observed to survive for over a year post-surgery without displaying behavioral differences compared to healthy, non-operated controls.

\textbf{The imaging system}. We use a wide-field mesoscope using off-the-shelf components. A 473 nm CW laser (MBL-III-473-100 mW, CNI) was expanded to 12 mm via a beam expander and a 4f lens system (f = 150 mm and 200 mm), then focused (f = 150 mm) and reflected by a 2-mm right-angle aluminum-coated microprism. The beam passed through a 2$\times$ /0.5 NA objective (MVPLAPO 2 XC, Olympus) to excite sample fluorophores. The microprism, positioned between the objective and camera, acted as a dichroic-like element, separating excitation and emission light~\cite{werley2017ultrawidefield}. Fluorescence was collected via the same objective, a tube lens (MVPLAPO 1X, Olympus), a dual-bandpass filter (520 $\pm$ 12.5 nm / 630 $\pm$ 46.5 nm, Edmund), and an sCMOS camera (ORCA-Flash4.0 V3, HAMAMATSU). The system provides a 6.6 mm FOV with a spatial resolution of 3.25 $\text{µm}$/pixel (2$\times$ magnification, 6.5 $\text{µm}$ pixel size).

\textbf{Behavioral paradigm}. To investigate the multiscale dynamics and causal mechanisms underlying consciousness, we induced three distinct stages—wakefulness, anesthesia, and post-anesthesia recovery—using the volatile anesthetics desflurane at 3\% (oxygen flow rate 0.5 L/min) or sevoflurane at 1.2\% (oxygen flow rate 0.5 L/min). Mice were head-fixed under the macroscope while allowed to move freely on a custom-built treadmill. After a 30-minute habituation period, imaging commenced at time zero with a 100-second awake baseline, followed by anesthesia administration from 100 to 700 seconds. Imaging continued throughout the recovery period until 1500 seconds.

\textbf{Fluorescence image processing and neuron extraction}. Given the high data throughput of the imaging system, we developed a parallelized analysis pipeline based on CNMF-E~\cite{kauvar2020cortical}. Raw videos were motion-corrected and temporally summarized into pixel-neighbor correlation and peak-to-noise ratio maps. Their Hadamard product initialized neuron candidates, which were refined via CNMF with a ring-shaped background model. Neurons overlapping with blood vessels were excluded using an intensity-based mask. Extracted temporal traces were further denoised using a supervised neural network trained to reject motion- and hemodynamics-related artifacts. Neuronal footprints were registered to the Allen CCF v3 using cranial window landmarks, enabling cortical area assignment~\cite{kauvar2020cortical}.

\subsection*{Data preprocessing}
\label{data_generation}

We utilized calcium signal data as model input, with the raw data comprising 15,000 time steps. To address the computational challenges posed by the high-dimensional raw data (including thousands of neurons), we first performed dimensionality reduction using K-means clustering, grouping the underlying neurons into 32 or 64 clusters based on their spatial coordinates where each cluster center is defined as the location of a ``functional neuronal ensemble'', then generated new neural signal data for each cluster by averaging calcium signal values of all neurons within each cluster while preserving the original temporal resolution. Moreover, since the data spans a wide range of values, we then standardized the data to facilitate model learning through Z-score.

Besides, since the data lengths of the three distinct stages (awake (0-1,000), anesthetized (1,000-7,000), and recovery (7,000-15,000)) differ significantly, we selected an equal number of time steps (1,000) from each stage as model inputs to ensure a fair comparison. The awake stage (0-1,000 time steps) required no modification, as it only contained 1,000 time steps. For the anesthetized and recovery stages, we selected the middle 1000 time steps from each stage as representative samples (4,000-5,000 and 10,000-11,000 time steps, respectively). This selection strategy ensures temporal consistency while maintaining state-specific characteristics.

\textbf{Generation of Shuffled Data:}
To assess temporal dependencies, we performed perturbation analysis by randomly permuting the dimensionality-reduced data along the temporal direction while maintaining equivalent time interval selection between shuffled and no-shuffled conditions. This randomization procedure was replicated 10 times to ensure statistical robustness.

\subsection*{Model} \label{model_nis+}

We utilize NIS+~\cite{yang2025finding}, a multiscale framework for quantifying the emergence of consciousness that maximizes Effective Information ($\mathcal{J}_d$) at each dimension while maintaining high predictive capability. The model analyzes time-series neural activity data from mouse brains, simultaneously quantifying causal effects across dimensions while learning emergent dynamics and optimal information aggregation strategies. Our approach consists of three distinct and progressive steps: (i) problem formulation, (ii) model architecture, and (iii) a two-stage training paradigm, each described in detail in the following sections.

\subsubsection*{Problem definition}

Consider a complex brain neural system where the neural data is represented as a $p$-dimensional time series $\{\boldsymbol{x}(t)\}$ for $t=1,2,...,\mathrm{T}$, forming observable micro-states. The objective is to maximize Effective Information ($\mathcal{J}_d$) of the macro-dynamics $f_d$ while ensuring the predicted state $\hat{\boldsymbol{x}}(t+1)$ remains $\epsilon-$close to the observed state $\boldsymbol{x}(t+1)$, defined as follows:

\begin{equation}
\begin{aligned}
\label{old_optimization}
&\max_{\phi_d,f_d,\phi^+_d} \mathcal{J}_d(f_d),\\
&s.t. \begin{cases}
|| \hat{\boldsymbol{x}}(t+1)-\boldsymbol{x}(t+1) || < \epsilon,\\
\hat{\boldsymbol{x}}(t+1)=\phi^{\dag}_d(f_d(\phi_d(\boldsymbol{x}(t)))).
\end{cases}
\end{aligned}
\end{equation}

Where $\epsilon$ denotes a predetermined small constant (Here, we set $\epsilon$=1 with reference to the result shown in Fig.~\ref{fig:dis_of_causal_power}g). $\phi_d : \mathcal{R}^p\rightarrow \mathcal{R}^d$ and $\phi^{\dag} _d
 :\mathcal{R}^d\rightarrow \mathcal{R}^p$ are the encoder (corresponding to the information aggregation strategy) and decoder function, respectively, where $d$ is the dimension of macro-states. The encoder comprises two operations: information transformation (implemented by invertible neural networks~\cite{dinh2016density}) and information discard; the decoder performs the inverse of the encoder and is responsible for information generation, where the discarded information is replaced by added Gaussian noise (for detailed definitions and notations, see reference~\cite{zhang2022neural}).

\textbf{Dimension-averaged effective information}

Since EI is susceptible to the influence of dimension $d$, which may complicate the comparison of EI across brain dynamical systems with different dimensions, it is important to emphasize that EI used here refers to dimension-averaged Effective Information (dEI). This metric objectively quantifies the strength of causal effects across varying dimensions, eliminates the bias introduced by dimensionality, and remains applicable to multivariate real-valued variables~\cite{zhang2022neural}.

 During our calculation, according to \cite{Liu2024exact}, dimension-averaged Effective Information ($\mathcal{J}_p$) can be decomposed into two parts, determinism and non-degeneracy, as
    \begin{eqnarray}\label{EIDeterminismDegeneracy}
    \begin{aligned}
    	\mathcal{J}_p
     &=\mathop{\ln\left[(2\pi e)^{-\frac{1}{2}}\det(\boldsymbol{\Sigma}_p^{-1})^\frac{1}{2p}\right]}_{Determinism}+\mathop{\ln\left[|\det(\nabla f_p(\boldsymbol{x}(t))|^\frac{1}{p}L\right]}_{Non-degeneracy},
     \end{aligned}
    \end{eqnarray}
where $\nabla f_p(\boldsymbol{x})$ is the gradient of dynamical function
    \begin{eqnarray}\label{dynamic}
    \begin{aligned}
 \boldsymbol{x}({t+1})=f_p(\boldsymbol{x}(t))+\boldsymbol{\varepsilon}_t, \boldsymbol{\varepsilon}_t\sim\mathcal{N}(0,\boldsymbol{\Sigma}_p)
     \end{aligned}
    \end{eqnarray}
in Eq.~(\ref{old_optimization}) at point $\boldsymbol{x}(t)$, where $\boldsymbol{x}(t)\in\mathcal{R}^p, \boldsymbol{\Sigma}_p\in\mathcal{R}^{p\times p}$. $f_p: \mathcal{R}^p\to \mathcal{R}^p$ captures deterministic dynamics, $\boldsymbol{\varepsilon}_t$ denotes random noise, accounting for perturbations effects, and $L$ represents the size of the probability space with a uniform distribution determined by the do-operator $do(\cdot)$, which is an intervention that enforces $\boldsymbol{x}(t)\sim U([-L, L]^p)$, where $U$ represents uniform distribution.

\subsubsection*{Causal emergence}

By varying the dimension $d$, we can derive macro-dynamics ($f_d$) at different dimensions. This allows for systematic comparison between $\mathcal{J}_d$ and $\mathcal{J}_p$ across all $d$. Formally, we quantify the degree of causal emergence by computing the following difference:

\begin{equation}
\label{eq:causal_emergence_identification}
\Delta\mathcal{J}_d\equiv \mathcal{J}_d-\mathcal{J}_p,
\end{equation}
Here, the micro-dynamics $f_p$ is derived by solving Eq.~(\ref{old_optimization}) with full dimensionality ($d=p$). For example, if $f_p$ and $f_d$ are the dynamical functions of micro and macro states, $\boldsymbol{\Sigma}_p$ and $\boldsymbol{\Sigma}_d$ are the covariance matrices of micro and macro states, $\Delta\mathcal{J}_d$ is expressed as
\begin{eqnarray}\label{Causal-emergence-baseline}
\Delta\mathcal{J}_d =\mathop{\ln\frac{|\det(\nabla f_d(\boldsymbol{y}(t))|^\frac{1}{d}}{|\det(\nabla f_p(\boldsymbol{x}(t))|^\frac{1}{p}}}_{Non-degeneracy\           Emergence}+\mathop{\ln\frac{|\det(\boldsymbol{\Sigma}_p)|^\frac{1}{2p}}{|\det(\boldsymbol{\Sigma}_d)|^\frac{1}{2d}}}_{Determinism\ Emergence}.
	\end{eqnarray}

\subsubsection*{Model architecture}

Solving the optimization problem defined in Eq.~(\ref{old_optimization}) directly proves challenging as the objective function $\mathcal{J}_d$ is a functional optimization problem.

To address this problem, we can mathematically prove that Eq.~(\ref{new_optimization})  can be derived from Eq.~(\ref{old_optimization}) by introducing inverse dynamics, inverse probability reweighting technique, and variational inequality~(detailed derivations are provided in reference~\cite{yang2025finding}), as follows:
\begin{gather}
\label{new_optimization}
\min_{f_d,g_d,\phi_d,\phi^{\dag}_d} \sum_{t=1}^{T-1} w(\boldsymbol{x}(t))||\boldsymbol{y}_d(t)-g_d(\boldsymbol{y}_d(t+1))||+\lambda||\hat{\boldsymbol{x}}(t+1)-\boldsymbol{x}(t+1) ||,
\end{gather}
where $\hat{\boldsymbol{x}}(t+1)=\phi^{\dag}_d(f_d[\phi_d(\boldsymbol{x}(t))])$. $\boldsymbol{y}_d(t)=\phi_d(\boldsymbol{x}(t))$ and $\boldsymbol{y}_d(t+1)=\phi_d(\boldsymbol{x}(t+1))$ are the macro-states at the $t$ and $(t+1)$ time step, respectively. $\phi_d$ is a encoder with hierarchical structures (see Appendix Fig.~\ref{fig:si_structure_encoders}). $g_d$ is a newly introduced function designed to simulate the inverse macro-dynamics for maximizing $\mathcal{J}_d$, mapping each macro-state at the $(t+1)$ time step back to its corresponding macro-state at the $t$ time step. $\lambda$ is a hyperparameter. $w(\boldsymbol{x}(t))$  denotes the inverse-probability weight that transforms the macro-state $\boldsymbol{y}_d(t)$ into a uniform distribution, expressed as $\frac{\Tilde{p}(\boldsymbol{y}_d(t))}{p(\boldsymbol{y}_d(t))}$, where $\Tilde{p}(\boldsymbol{y}_d(t))$ is equal to uniform distribution and $p(\boldsymbol{y}_d(t))$ represents the probability distribution of the macro-states that is approximated using
kernel density estimation~\cite{rosenblat1956remarks}.

\begin{figure}[!htp]
	\centering
	\includegraphics[width=0.75\linewidth]{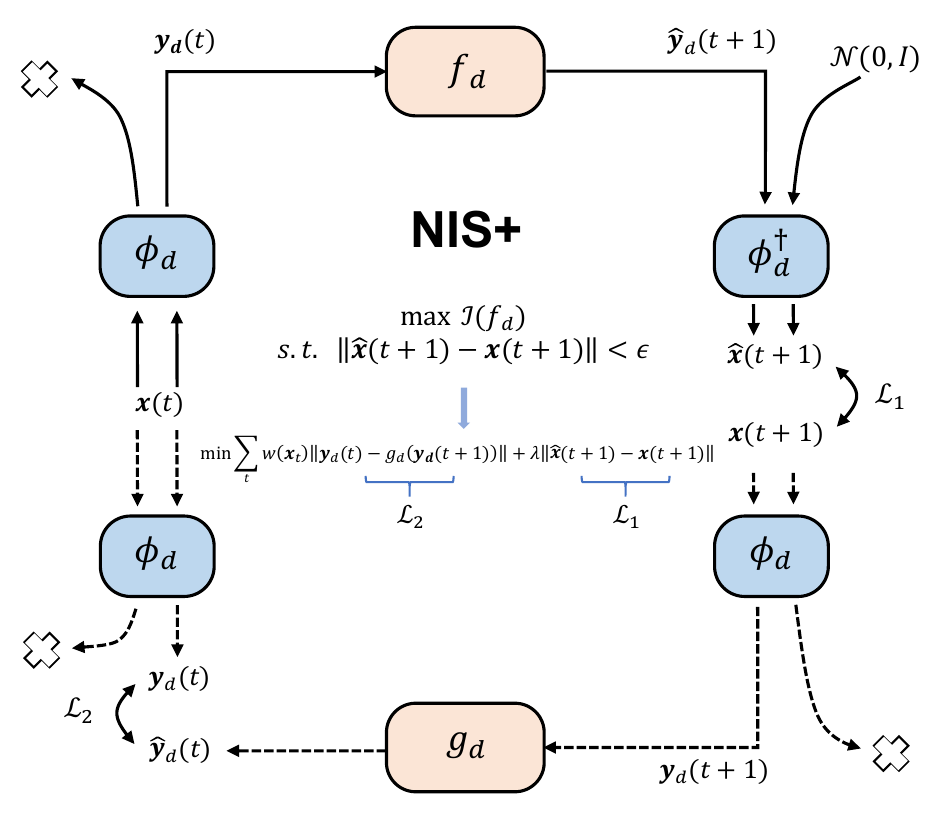}
	\caption{\textbf{The architecture of NIS+ in one of the dimensions ($d$)}.  $\boldsymbol{x}(t)$ and $\boldsymbol{x}(t+1)$ represent the micro-states, while $\hat{\boldsymbol{x}}(t+1)$ represents the predicted micro-state. $\boldsymbol{y}_d(t)=\phi_d(\boldsymbol{x}(t))$ and $\boldsymbol{y}_d(t+1)=\phi_d(\boldsymbol{x}(t+1))$ represent the macro-states obtained by encoding the micro-states at the time steps $t$ and $(t+1)$, respectively. ${\phi}_d$ and ${\phi}^{\dag}_d$ represent the encoder and decoder, respectively. The encoder (${\phi}_d$) consists of two operations: information transformation (achieved via invertible neural networks~\cite{dinh2016density}) and information discarding, while the decoder~(${\phi}^{\dag}_d$) is the inverse process of the encoder, where information discarding is replaced by adding Gaussian noise~($\mathcal{N}(0, I)$). $\hat{\boldsymbol{y}}_t$ and $\hat{\boldsymbol{y}}_{t+1}$ represent the predicted macro-states by inverse~($g_d$) and forward dynamics~($f_d$), respectively. The objective function is designed to simultaneously optimize forward error ($L_1=\hat{\boldsymbol{x}}(t+1)-\boldsymbol{x}(t+1)$) and backward error ($L_2=g_d(\boldsymbol{y}_d(t+1))-\boldsymbol{y}_d(t)$). $\lambda$ is a weighting factor combining $L_1$ and $L_2$. $w(\boldsymbol{x}_t)$ represents inverse probability weights,  which is equal to $\frac{\Tilde{p}(\boldsymbol{y}_d(t))}{p(\boldsymbol{y}_d(t))}$. Here, $\Tilde{p}(\boldsymbol{y}_d(t))$ denotes the modified probability distribution of macro-state $\boldsymbol{y}_d(t)$ after the intervention $do$($\boldsymbol{y}_d(t) \sim U_d$), while $p(\boldsymbol{y}_d(t))$ represents the probability distribution of the macro-states that can be approximated using
kernel density estimation~\cite{rosenblat1956remarks}. The boxes represent neural networks, and arrows pointing to a cross denote information discarding operations.} 
	\label{fig:nis+}
\end{figure}

To solve the converted problem defined in Eq.~(\ref{new_optimization}), we design a machine learning framework (NIS+) as shown in Fig.~\ref{fig:nis+}. The model contains two information-flow pathways:
(1) encoder($\phi_d$) $\rightarrow$ forward dynamic($f_d$) $\rightarrow$ decoder($\phi^{\dag}_d$) (solid-line); (2) encoder($\phi_d$) $\rightarrow$  inverse dynamic($g_d$) (dashed-line). From these, we obtain a forward error  ($L_1=\hat{\boldsymbol{x}}(t+1)-\boldsymbol{x}(t+1)$) and an inverse error ($L_2=g_d(\boldsymbol{y}_d(t+1))-\boldsymbol{y}_d(t)$). A total loss function is formed by combining $L_1$ and $L_2$ with a weighting factor $\lambda$ ($\lambda=1$ for the fair treatment of these two errors), and the overall objective~(Eq.~(\ref{new_optimization})) is optimized via gradient updates.

\subsubsection*{Training methodology}

To optimize the model, we adopt a two-stage training paradigm here. The first stage trains exclusively the forward dynamics (upper information channel in  Fig.~\ref{fig:nis+}) to minimize prediction error as a pretraining process, while the second stage jointly optimizes both forward and inverse dynamics (upper and lower channels in  Fig.~\ref{fig:nis+}) to maximize dimension-averaged Effective Information~($\mathcal{J}_d$).

Fig.~\ref{fig:nis+} only depicts a single-scale model in dimension $d$; in practice, to enhance computational efficiency, we adopt a multistory framework, as shown in reference~\cite{yang2025finding}, with two key modifications during the training process: (i) parallel training of all forward dynamics across scales during stage one, using the average microscopic loss of all scales for gradient updates; and (ii) maximizing ($\mathcal{J}_d$) independently for each scale, as shown in Fig.~\ref{fig:dis_of_causal_power}d. The detailed parameter settings of the model can be found in appendix~\ref{model_parameter}.

Additionally, it should be noted that for stage two, there are two training approaches of NIS+: one is the method adopted in this paper, which independently maximizes $\mathcal{J}_{d=2^k}$ at each level $k$; the other involves fixing the encoder parameters below the $k^{th}$ layer and only training the current layer $\tilde{\phi}_{2^k}$ while maximizing $\mathcal{J}_{d=2^k}$. In Appendix~\ref{sec:si_compare_two_train_mehtod}, we provide a detailed theoretical and experimental comparison of these two approaches, demonstrating that the two training methods are approximately equivalent.

\subsection*{Baseline} 
\label{baseline}

When the baseline represents the results obtained when the data exhibits Brownian motion dynamics, the dynamical function trained by NIS+ satisfies $f_p(\boldsymbol{x}(t))=\boldsymbol{x}(t)\in\mathcal{R}^p$ (corresponding to the dynamical function in Eq.~(\ref{dynamic})), indicating that we only study the impact of noise on the system and the variables remain constant over time if $\boldsymbol{\Sigma}_p\to 0$. We can also get $\nabla f_p(\boldsymbol{x}(t))=\boldsymbol{I}_p\in\mathcal{R}^{p\times p}$, $f_d(\boldsymbol{y}(t))=\boldsymbol{y}(t)\in\mathcal{R}^d$, and $\nabla f_d(\boldsymbol{y}(t))=\boldsymbol{I}_d\in\mathcal{R}^{d\times d}$. Then the Non-degeneracy emergence in Eq.~(\ref{Causal-emergence-baseline}) satisfy
\begin{eqnarray}\label{nondegeneracy-baseline}
\ln\frac{|\det(\nabla f_d(\boldsymbol{y}(t)))|^\frac{1}{d}}{|\det(\nabla f_p(\boldsymbol{x}(t)))|^\frac{1}{p}}=0,
\end{eqnarray}
which means only the covariance of micro and macro noise is calculated, and the dynamical functions $f_p$ and $f_d$ of the system are not considered as
\begin{eqnarray}\label{Causal-emergence-baseline-brown}
\Delta\mathcal{J}_d =\mathop{\ln\frac{|\det(\boldsymbol{\Sigma}_p)|^\frac{1}{2p}}{|\det(\boldsymbol{\Sigma}_d)|^\frac{1}{2d}}}
	\end{eqnarray}
is only determined by covariance matrices $\boldsymbol{\Sigma}_p$ and $\boldsymbol{\Sigma}_d$ of micro and macro dynamics, which reflect the difference between the determinism of the two dynamics.

\subsection*{The calculation of coverage} \label{cal_coverage}

Next, we detail the method for calculating the coverage of a macro-variable in Fig.~\ref{fig:information_integration}b. This coverage describes the spatial distribution extent of the micro-variables significantly associated with each macroscopic causal variable. The core idea of the method is to transform the attribution relationships and strengths between variables at different levels into a multi-layer Markov process, analogous to a Galton board~\cite{galton1889natural} where a ball falls from the top layer by layer. Starting from the top-level variable and attributing downward is like releasing a rolling ball: the ball falling layer by layer corresponds to the influence range formed by the layer-by-layer attribution of higher-level variables to lower-level variables. The final stable distribution of the ball at the bottom layer analogously represents the stable attribution probability of the top-level variable over the bottom-level variables. Based on this probability distribution, the coverage of the top-level variable relative to the bottom layer can be computed.

\begin{figure}[!htp]
	\centering
	\includegraphics[width=0.79\linewidth]{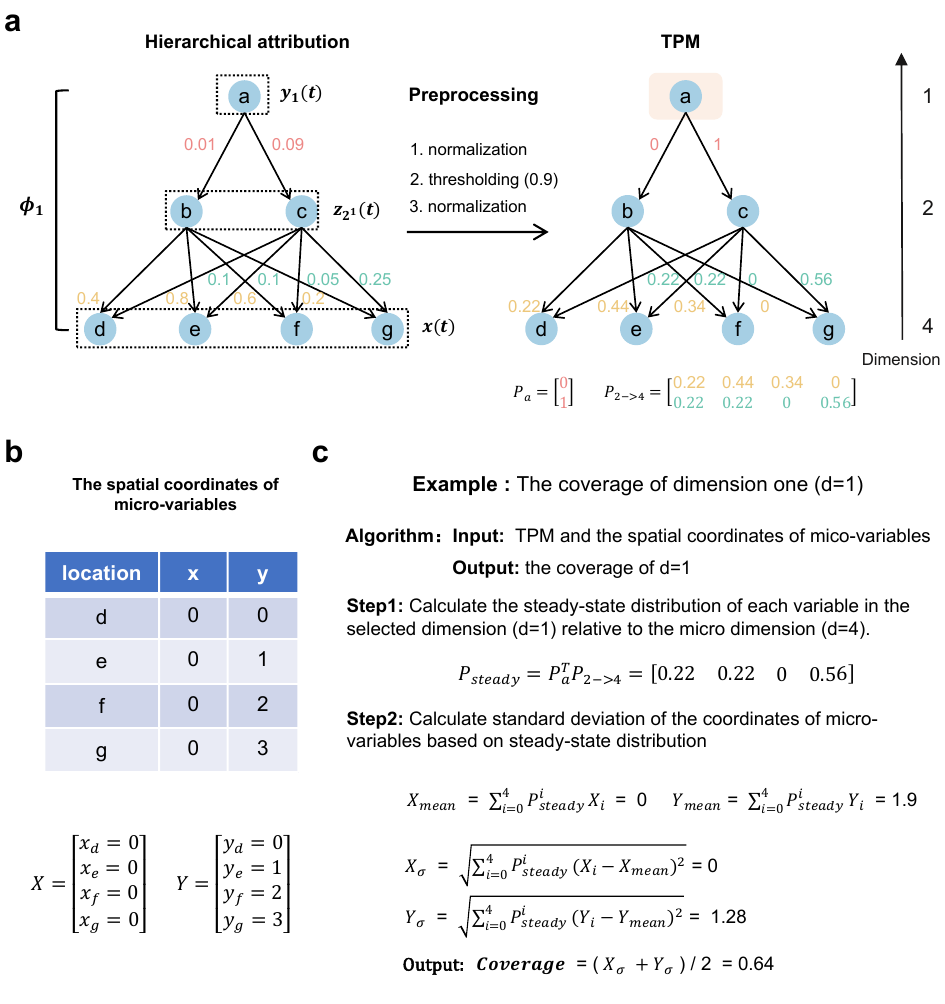}
	\caption{The detailed calculation process of the spatial distribution extent (``coverage”) of micro-variables significantly associated with each macroscopic causal variable in Fig. \ref{fig:information_integration}b. This algorithm computes the coverage corresponding to each macro-variable based on the attribution relationships from high- to low-level variables and the coordinate positions of the bottom-level variables. \textbf{a} The illustration of multiscale attribution and probability transition graphs. To simplify the explanation, we assume a hierarchical three-layered neural network encoder $\phi_1$ and seven variables representing causal variables at different scales.  Among them, $d$, $e$, $f$, and $g$ reside at the third layer, representing four input variables (i.e., neuronal functional ensembles), which constitute $\boldsymbol{x}(t)$; $b$ and $c$ are at the second layer, forming a two-dimensional intermediate variable $\boldsymbol{z}_2(t)$; and $a$ is at the first layer, constituting the one-dimensional top-level macro-variable $\boldsymbol{y}_1(t)$. Left: Directed multiscale attribution graph. Edges represent attribution relationships between variables in adjacent layers, with weights indicating the magnitude of attribution values. These values are obtained by computing integrated gradients of the output of the $i^{th}$ layer encoder with respect to its inputs, and then averaging across multiple samples. For simplicity, a set of hypothetical attribution values is illustrated. Right: Transition probability graph between variables at different layers. The transition probability matrix (TPM) is derived from the left attribution graph through preprocessing, which includes three steps: normalization, thresholding, and renormalization. Normalization converts the attribution values of connections from layer $i$ to layer ($i+1$) into transition probabilities. Thresholding filters edges with low probability by setting transition probabilities below 0.9 to zero. Renormalization ensures the thresholded TPM satisfies the normalization condition.  \textbf{b} Assumed two-dimensional coordinates of each micro-variable. \textbf{c} The detailed calculation process of the ``coverage” for variable $a$ $(d=1)$. The algorithm takes the TPM and the coordinates of micro-variables as inputs and outputs the coverage value of $a$. The calculation involves two steps: 1) Compute the steady-state distribution of the macroscopic causal variable $\boldsymbol{y}_1(t)$  transferred layer by layer through the TPM to the micro-variables $\boldsymbol{x}(t)$ and take it as the probability distribution of the degree of dependence of this macroscopic causal variables on the micro-variables; 2) Calculate the standard deviation of the coordinates of the micro-variables based on the distribution from step one as the coverage of the macroscopic causal variable $\boldsymbol{y}_1(t)$.}
	\label{fig:calculation_coverage}
\end{figure}

To illustrate this calculation method, we assume that the encoder in NIS+ consists of only three layers (Fig.~\ref{fig:calculation_coverage}a), corresponding to dimensions 1, 2, and 4. The variables at each scale are denoted as $\boldsymbol{y}_1(t)$ (including the top-level variable $a$), $\boldsymbol{z}_2(t)$  (including the intermediate-level variables $b$, $c$, $d$), and $\boldsymbol{x}(t)$  (including the bottom-level variables $e$, $f$, $g$, $h$), where $\boldsymbol{x}(t)$ represents micro-variables and the others represent macro-variables.

Fig.~\ref{fig:calculation_coverage}a (left) shows the multiscale attribution graph, depicting the dependency relationships and attribution strengths between variables at different scales. The connection weights indicate the degree of attributional dependence of causal variables at layer $i$ on variables at layer ($i+1$) obtained via integrated gradient algorithms~\cite{sundararajan2017axiomatic}. For simplicity, a set of hypothetical attribution values is assumed (left panel). Through preprocessing steps (including normalization, thresholding, and renormalization), these attribution relationships are transformed into a transition probability matrix (TPM) (right panel). To compute the coverage, we assume a set of two-dimensional coordinates in the brain for each micro-variable, as shown in Fig.~\ref{fig:calculation_coverage}b. Fig.~\ref{fig:calculation_coverage}c illustrates the calculation process for the coverage of the one-dimensional variable $a$. The inputs required for the computation are: the TPM between variables and the two-dimensional coordinates of each micro-variable. The final output is the coverage of variable $a$. The algorithm consists of two computational steps:

\begin{itemize}
    \item Compute the steady-state distribution ($\boldsymbol{P}_{steady}$) of variable $a$ in the target dimension ($d$=1) as it transfers layer by layer through the TPM to the micro-dimension ($d$=4). This distribution represents the probabilistic dependence of variable $a$ on each variable at the micro-scale, where $\boldsymbol{P}_{steady}=\boldsymbol{P}^T_a \cdot \boldsymbol{P}_{2 \rightarrow 4}$. Here, $\boldsymbol{P}_a$ denotes the transition probability vector of variable $a$ and $\boldsymbol{P}_{2 \rightarrow 4}$ represents the transition probability matrix from variables at dimension 2 ($d$=2, i.e., $b$, $c$) to variables at dimension 4 ($d$=4, i.e., $d$, $e$, $f$, $g$).

    \item Based on the steady-state distribution ($\boldsymbol{P}_{steady}$) obtained in step one, calculate the standard deviations ($\boldsymbol{X}_\sigma$ and $\boldsymbol{Y}_\sigma$) of the coordinates of the random variable  $\boldsymbol{x}(t)$, and take the average of the two-dimensional standard deviations $\boldsymbol{X}_\sigma$ and $\boldsymbol{Y}_\sigma$ as the ``coverage" of the variable $\boldsymbol{y}_1(t)$ at dimension $d=1$.

\end{itemize}

In more general cases where the top level contains multiple variables, the coverage is computed individually for each variable, and the final coverage of the top-level variables is obtained by averaging these values.

\section*{Data Availability}
All data used in this study can be accessed at our GitHub repository: \newline https://github.com/

\section*{Code Availability}
All code used in this study can be accessed at our GitHub repository: \newline https://github.com/

\section*{Acknowledgments}
The authors would like to acknowledge all the members who participated in the "Causal Emergence Reading Group" organized by the Swarma Research and the "Swarma - Kaifeng" Workshop. Besides, we are especially grateful to Professors Erik Hoel, Yuhan Chen and Dr Lingbo Li for their invaluable contributions and insightful discussions throughout the research process. We also thank the large language model DeepSeek for its assistance in refining our paper writing.

\section*{Author Contributions Statement}
J.Z and J.H conceived and directed the project. Z.P.W, J.H, and J.Z designed the model. Z.P.W conducted the experiments.  All the participants wrote the paper. All authors discussed the results and commented on the manuscript.

\section*{Competing Interests Statement}
The authors declare no competing interests.

\begin{appendices}

\section{The detailed structure of encoders}\label{sec:si_structure_encoders}

\begin{figure}[!ht]
    \centering
    \includegraphics[width=0.75\textwidth]{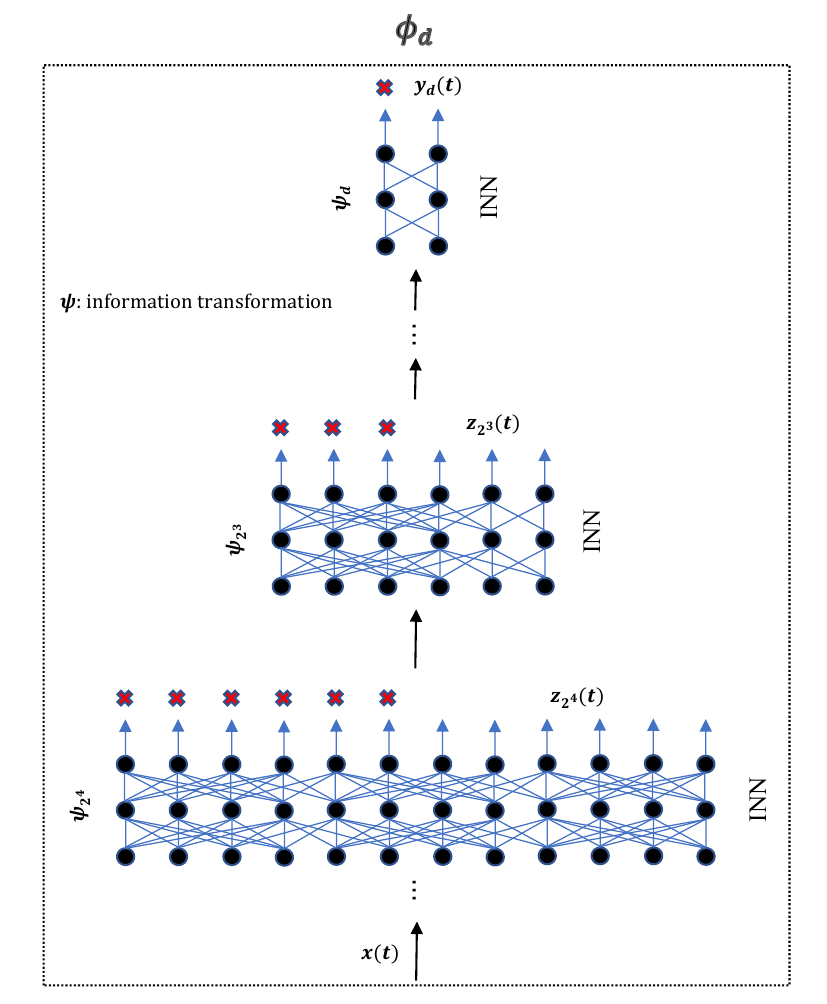}
    \caption{\textbf{The detailed hierarchical structure of the encoder $\phi_d$.} The encoder $\phi_d$ is a hierarchical structure composed of multiple invertible network layers, where each layer performs a lossless information transformation ($\psi_{2^{i}}$) and reduces the dimensionality by half (except for $d=5$). The information transformation is implemented by a multilayer invertible neural network (RealNVP~\cite{dinh2016density}). The encoder $\phi_d$ takes the raw data $\boldsymbol{x}(t)$ as input, and through layer-by-layer propagation, intermediate variables $\boldsymbol{z}_{2^i}(t)$ at different levels $i$ are obtained, ultimately forming the macroscopic causal variable $\boldsymbol{y}_{d}(t)$ in dimension $d$.}
    \label{fig:si_structure_encoders}
\end{figure}

\section{The comparison of two training methods for maximizing dEI}\label{sec:si_compare_two_train_mehtod}

\begin{figure}[!ht]
    \centering
    \includegraphics[width=0.9\textwidth]{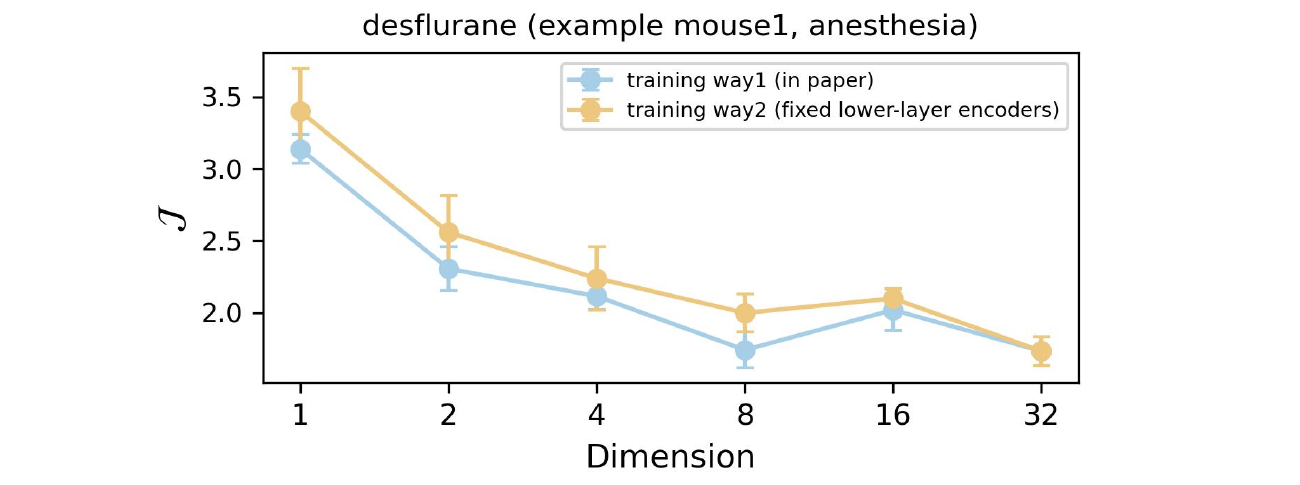}
    \caption{The comparison of $\mathcal{J}$ for two training approaches across different dimensions in mouse1 (deflurane), specifically including: the method of independently maximizing dimension-averaged Effective Information at each level $k$ employed in this paper (blue), and the approach of fixing the encoder parameters below the $k^{th}$ layer while training only the current layer ($\tilde{\phi}_{2^k}$) and maximizing dimension-averaged Effective Information (yellow). The results shown here are under anesthesia. These results represent the average of three repeated experiments, with error bars indicating the standard deviation.}
\label{fig:si_compare_two_train_method}
\end{figure}

NIS+ has two training approaches: The first, as adopted in this paper, involves independently maximizing dimension-averaged Effective Information at each layer $k$ (Fig.~\ref{fig:schematic_diagram}d). This method allows independent training at each scale. The second approach starts from layer $k=5$, progressively decreases $k$, and obtains $\boldsymbol{y}_{2^k}(t)$ by training through the maximization of Eq.~(\ref{new_optimization}). Since the encoder has $k$ layers for each dimension $d=2^k$, this method fixes the parameters of the encoder layers $i \textgreater k$, trains only the $k^{th}$ layer encoder $\tilde{\phi}_{2^k}$, and maximizes the dimension-averaged Effective Information $\mathcal{J}_{d={2^k}}$. This approach requires training lower scales before higher scales. Upon completion of training, intermediate variables $\boldsymbol{z}_{2^i}(t)$ ($i \in \left\{k+1, ..., 5\right\}$) at different levels can be obtained. It is noteworthy that these two training approaches are approximately equivalent from the perspective of the final multi-scale causal power.

To verify the approximate equivalence of the two training methods, we compared the dimension-averaged Effective Information ($\mathcal{J}_{d}$) for both approaches across different dimensions in mouse1, as shown in Fig.~\ref{fig:si_compare_two_train_method}. The results indicate that the trends of $\mathcal{J}_{d}$ with respect to dimensionality are largely consistent between the two methods: both exhibit an overall increasing trend as dimensionality decreases, with the strongest causal effects observed at the top macro-scale. Additionally, the orange curve generally lies above the blue curve. This occurs because the training method with fixed lower-layer parameters only optimizes the encoder at the current layer, involving very few tunable parameters. As a result, the overall prediction difficulty increases, leading to higher prediction errors. According to Eq.~(\ref{old_optimization}), the prediction error acts as a constraint. Therefore, when the achievable error of the model increases, the constraint force correspondingly decreases, allowing a higher level of maximized Effective Information under looser restrictions.

Furthermore, we attempted to increase the weight $\lambda$ in Eq.~(\ref{new_optimization}) to further reduce the micro-scale prediction error. However, the error remained slightly higher than that achieved by the training method used in this study. Despite this, the trends of $\mathcal{J}_{d}$ obtained from both methods remain consistent, and the overall $\mathcal{J}_{d}$ across different scales are approximately equal, with differences of less than 0.4 at each dimension. Thus, we can indirectly conclude that the two training methods are functionally equivalent.

Additionally, an important conclusion to supplement is that the intermediate variables $\boldsymbol{z}_{2^i}(t)$, obtained by maximizing dimension-averaged Effective Information at a given scale $d=2^k$, are approximately equal to the causal variables $\boldsymbol{y}_{2^i}(t)$ at dimensions smaller than $d$, i.e., $\boldsymbol{z}_{2^i}(t) \approx \boldsymbol{y}_{2^i}(t)$ for $i \in \left\{k+1,\dots,5 \right\}$. This conclusion can be justified as follows. First, since the two training approaches of NIS+ yield approximately equivalent results, and in the second approach (with fixed lower-layer parameters), we maximize dEI for variables of dimension $2^k$ while progressively decreasing $k$, the resulting intermediate variables $\boldsymbol{z}_{2^i}(t)$ (for any $5 \textgreater i \textgreater k$) are essentially equal to $\boldsymbol{y}_{2^i}(t)$. Given the equivalence between the two training methods, it follows that the hierarchical variables $\boldsymbol{z}_{d}(t)$ obtained via the first method are also approximately equal to $\boldsymbol{y}_{d}(t)$. This conclusion can also be explained from another perspective. According to the findings in the work on invertibility~\cite{zhang2025dynamical}, when maximizing dEI at the $k^{th}$ level, the optimal encoder effectively projects the raw data onto the subspace spanned by the singular vectors corresponding to the largest $2^k$ singular values. Similarly, when maximizing dEI at the $(k+1)^{th}$ level, the optimal encoder projects the data onto the subspace spanned by the top $2^{k+1}$ singular vectors. Since the subspace spanned by the top $2^{k}$  singular vectors is a subset of that spanned by the top $2^{k+1}$ singular vectors, the first $(5-k)$ layers of the encoder optimized for $\boldsymbol{y}_{2^k}(t)$ are approximately identical to those obtained when maximizing $\boldsymbol{y}_{2^{k+1}}(t)$ under optimal conditions.

Therefore, although the method described in the main text yields a set of mutually independent variables $\boldsymbol{y}_d(t)$, these variables can approximately be viewed as forming a hierarchical structure, similar to the hierarchical intermediate variables $\boldsymbol{z}_d(t)$ generated when optimizing variables $\boldsymbol{y}_d(t)$. This also explains why the one-dimensional dynamics in Fig.~\ref{fig:dynamics_analyse} of the main text can be interpreted as an approximate projection of the two-dimensional dynamical vector field.

\section{The detailed calculation process of the projection}\label{sec:si_cal_projection}

The projection is a straight line passing through the origin. $\tilde{\phi'}_1$ refers to the projection relationship between two macro-dynamics, obtained by calculating the Jacobian matrix of the encoder between the two macro-scales. Specifically, each macro-state $\boldsymbol{y}_{2}(t)$ is input into the encoder $\phi_1$ to obtain the macro-state $\boldsymbol{y}_{1}(t)$. Then, the derivative of $\boldsymbol{y}_{1}(t)$ with respect to $\boldsymbol{y}_{2}(t)$ is calculated to obtain the corresponding Jacobian matrix. By averaging the Jacobian matrices across all states, the final projection direction $\tilde{\phi'}_1$ can be determined, where $\tilde{\phi'}_1(\tilde{\phi'}^1_1,\tilde{\phi'}^2_1)$ is a two-dimensional vector. The values in this two-dimensional vector are then normalized, and the ratio of the two normalized directions is taken as the final slope of the projection: $\frac{\tilde{\phi'}^2_1}{\tilde{\phi'}^1_1}$.

\section{Robustness Analysis}\label{sec:si_robustness}

\begin{figure}[!ht]
    \centering    \label{fig:si_robustness_temporal_coarse_graining}
    \includegraphics[width=0.9\textwidth]{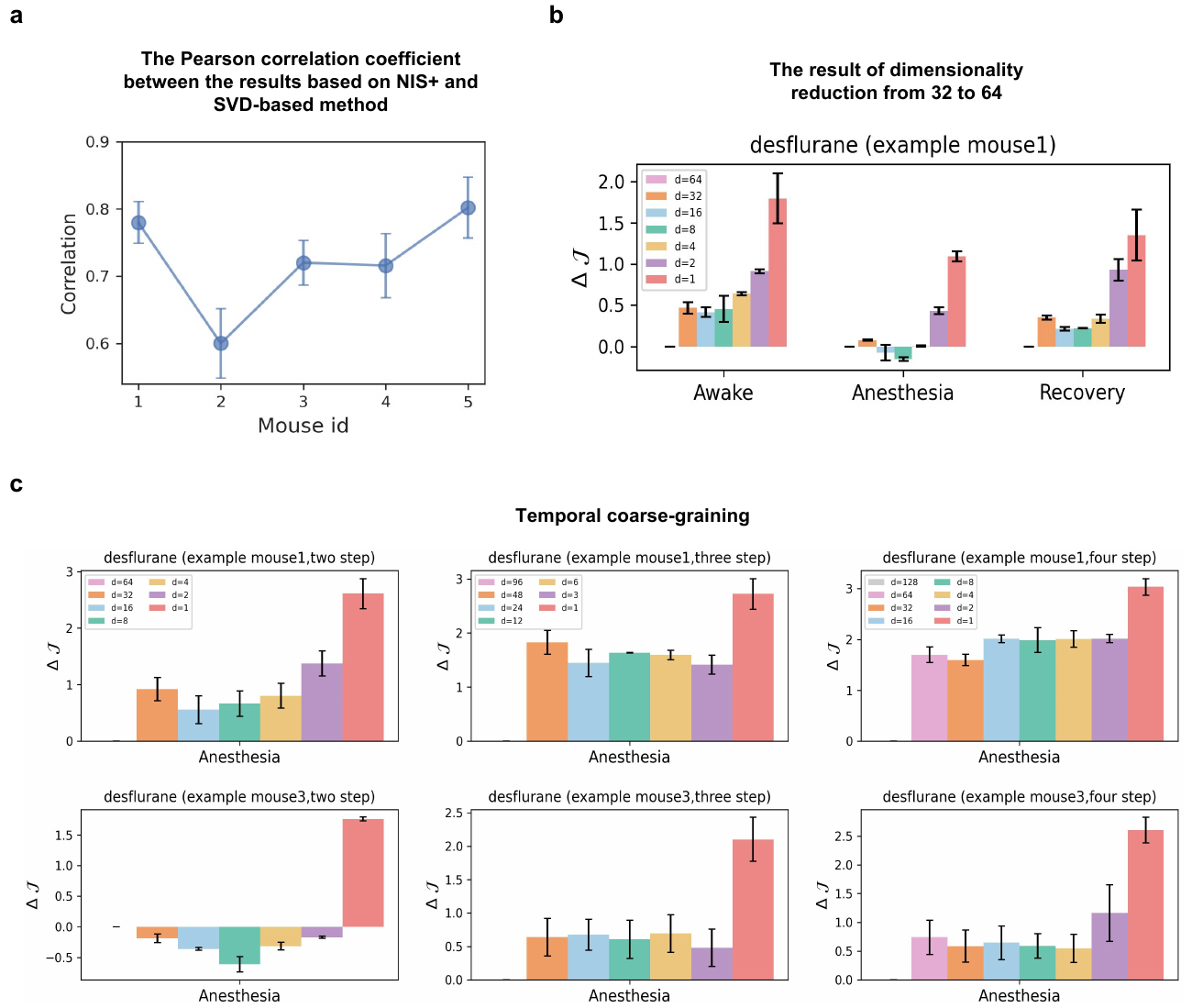}
    \caption{\textbf{Robustness analysis under desflurane}. \textbf{a} The Pearson correlation coefficient between $\Delta \mathcal{J}_1$ and  $\gamma_1$ calculated from NIS+ and SVD-based method in dimension one ($d$=1), respectively. \textbf{b} The comparison of causal emergence~($\Delta \mathcal{J}_d$)  across different stages and dimensions in mouse1. Here, the input data for the model is reduced to 64 dimensions.  \textbf{c} Temporal coarse-graining analysis of $\Delta \mathcal{J}_d$ across dimensions in mouse1 and mouse3, comparing two-, three-, and four-step predictions (e.g., two-step: $\boldsymbol{x}_t$,$\boldsymbol{x}_{t+1}$ $\to$ $\boldsymbol{x}_{t+2}$,$\boldsymbol{x}_{t+3}$). All analyses in Fig.~\ref{fig:si_robustness_temporal_coarse_graining}a and Fig.~\ref{fig:si_robustness_temporal_coarse_graining}c used 32-dimensional reduced data. The results are the average of three repeated experiments, and the error bar represents the standard deviation.}

\end{figure}

To further validated the robustness of our model, we propose two supplementary experiments: (1) comparison with an SVD-based causal emergence identification method~\cite{liu2025svd}, revealing strong correlation (Pearson's r $\textgreater$ 0.5) between $\Delta \mathcal{J}_1$ and  $\gamma_1$  on dimension one ($d$=1) across all mice (Fig.~\ref{fig:si_robustness_temporal_coarse_graining}a); and (2) temporal coarse-graining with multiple input/output time steps (Fig.~\ref{fig:si_robustness_temporal_coarse_graining}c), which consistently observed that dimension 1 exhibits the largest $\Delta \mathcal{J}_1$.

To assess the consistency of our qualitative conclusions with increasing dimensionality, we extended the K-means dimensionality reduction from 32 to 64 dimensions. As demonstrated in Fig.~\ref{fig:si_robustness_temporal_coarse_graining}b, dimension one ($d$=1) maintained the largest $\Delta \mathcal{J}_1$, with the relative ordering of $\Delta \mathcal{J}_1$ preserved across three stages~(awake $\textgreater$ recovery $\textgreater$ anesthesia), confirming the robustness of our findings.

\section{The comparison of model prediction errors}\label{sec:si_error}

\begin{figure}[!ht]
    \centering
    \includegraphics[width=0.8\textwidth]{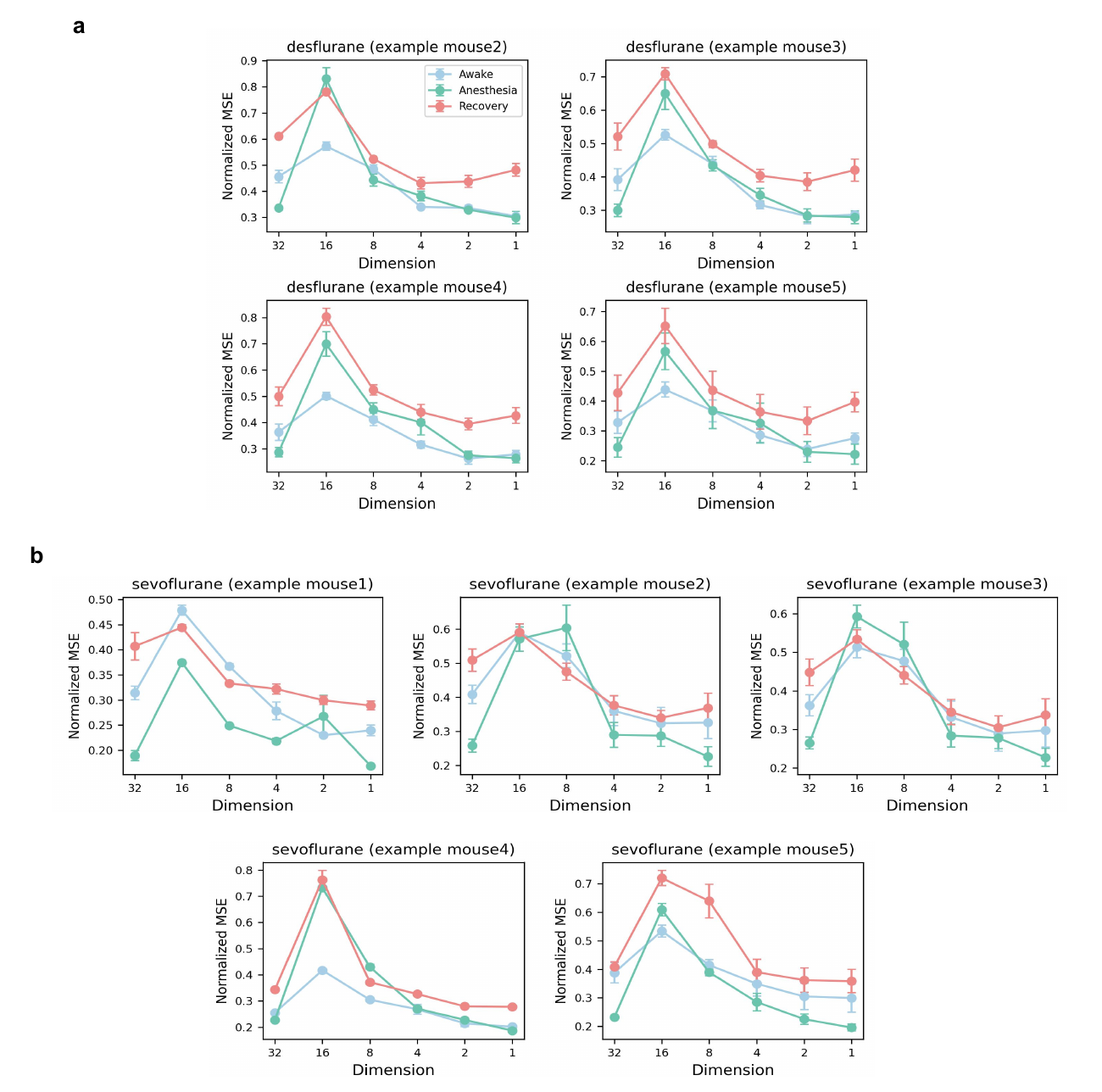}
    \caption{\textbf{Normalized MSE of the NIS+ model under the test set at different dimensions in three different stages, which is obtained by dividing the predicted MSE by the data variance.} \textbf{a} The results in desflurane from mouse2 to mouse5. \textbf{b} The results in sevoflurane from mouse1 to mouse5. The results are the average of three repeated experiments, and the error bar represents the standard deviation.}
    \label{fig:si_normalized_mse}
\end{figure}

\begin{figure}[!ht]
    \centering
    \includegraphics[width=0.8\textwidth]{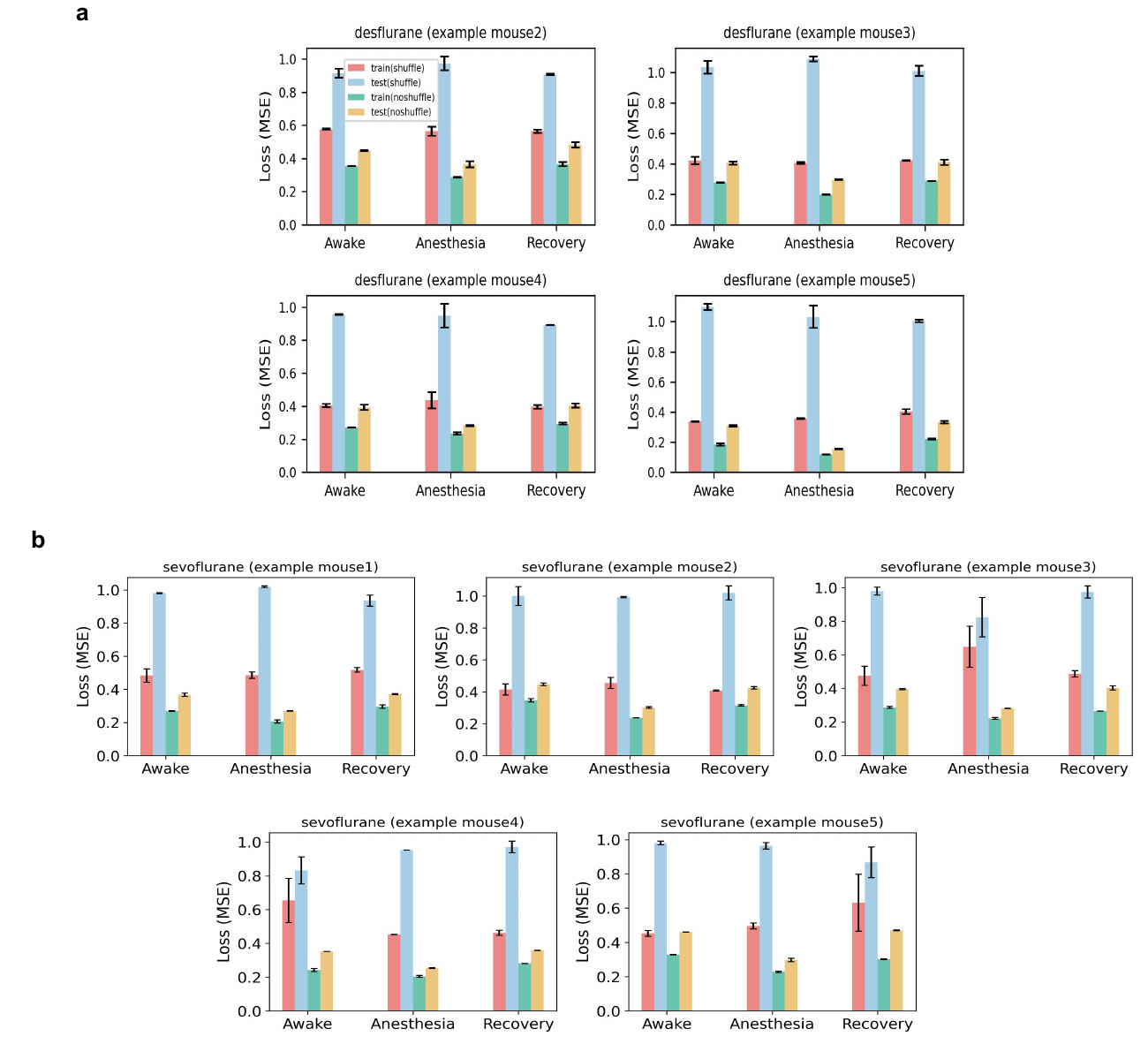}
    \caption{\textbf{The comparison of training and test errors of shuffled or not in different stages in dimension one ($d=1$).} \textbf{a} The results in desflurane from mouse2 to mouse5. \textbf{b} The results in sevoflurane from mouse1 to mouse5. The results are the average of three repeated experiments for the original data and ten repetitions for the shuffled data, and the error bar represents the standard deviation.}
    \label{fig:si_train_test_error}
\end{figure}

To evaluate the predictive performance of the multiscale dynamical model, we calculated the Normalized MSE (NMSE) across all dimensions and compared the errors before and after data shuffling (Fig.~\ref{fig:si_normalized_mse}-\ref{fig:si_train_test_error}). The results show that: (1) All NMSE values are below 1, indicating that the prediction error is lower than the data variance, which confirms the model's predictive capability~(Fig.~\ref{fig:si_normalized_mse}); (2) The error for the original (no-shuffled) data is significantly lower than that for the shuffled data, with a small gap between training and testing errors, demonstrating that the model avoids overfitting while capturing meaningful neural activity patterns~(Fig.~\ref{fig:si_train_test_error}). The conclusion remains consistent across different mice and under different anesthetics.

\section{The comparison of causal power and causal emergence across different stages}\label{sec:si_ce}

\begin{figure}[!ht]
    \centering
    \includegraphics[width=0.8\textwidth]{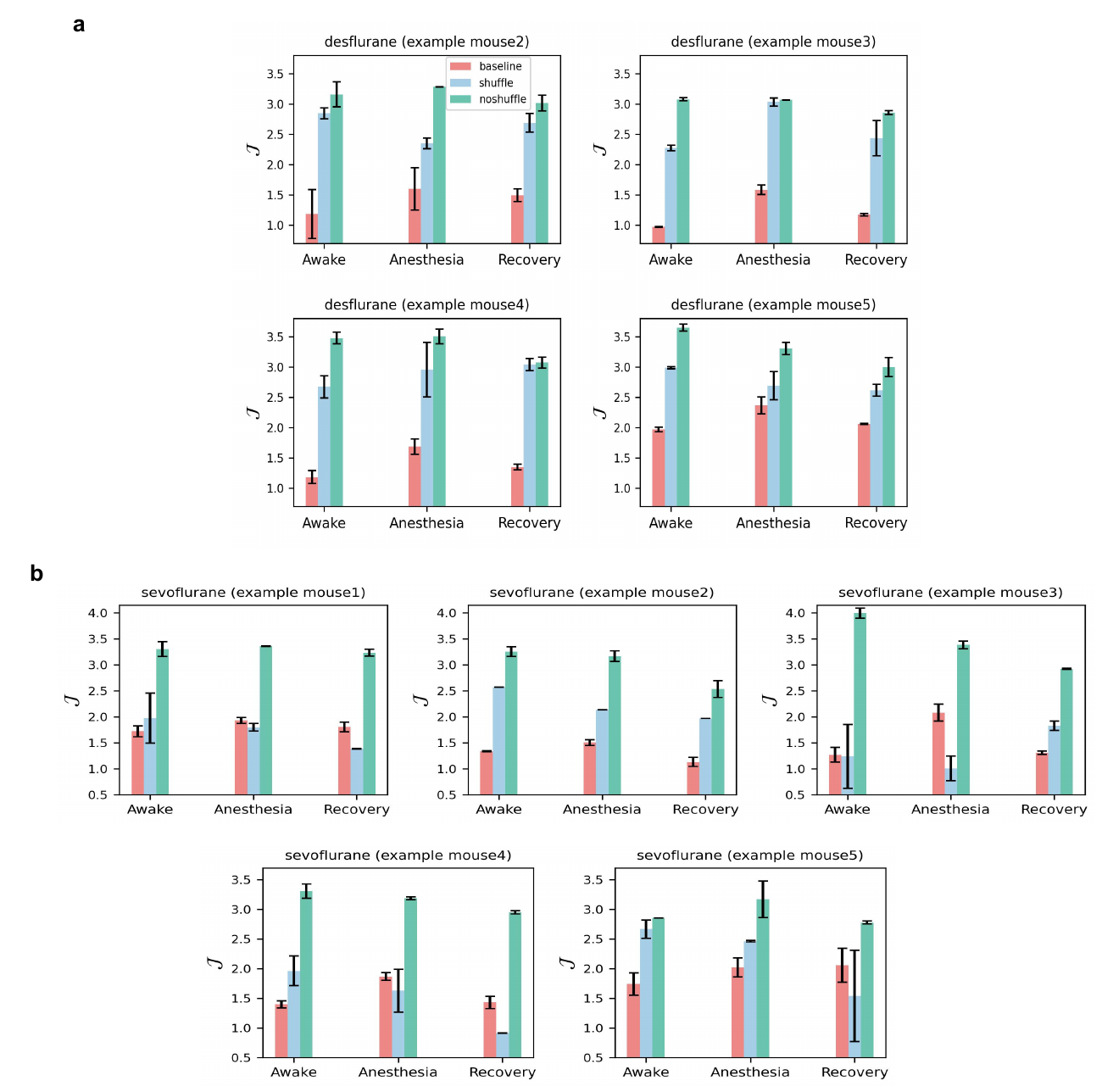}
    \caption{The comparison of the strength of Effective Information ($\mathcal{J}_1$) between shuffled/non-shuffled
conditions and baseline across three stages in $d$=1, where “shuffle” indicates random temporal permutation of the data. Here, “baseline” represents the results when the mice exhibit Brownian motion dynamics. “average” represents the mean results of the five mice. \textbf{a} The results in desflurane from mouse2 to mouse5. \textbf{b} The results in sevoflurane from mouse1 to mouse5. The results are the average of three repeated experiments for the original data and ten repetitions for the shuffled data, and the error bar represents the standard deviation.}
    \label{fig:si_ei_different_state}
\end{figure}

\begin{figure}[!ht]
    \centering
    \includegraphics[width=0.8\textwidth]{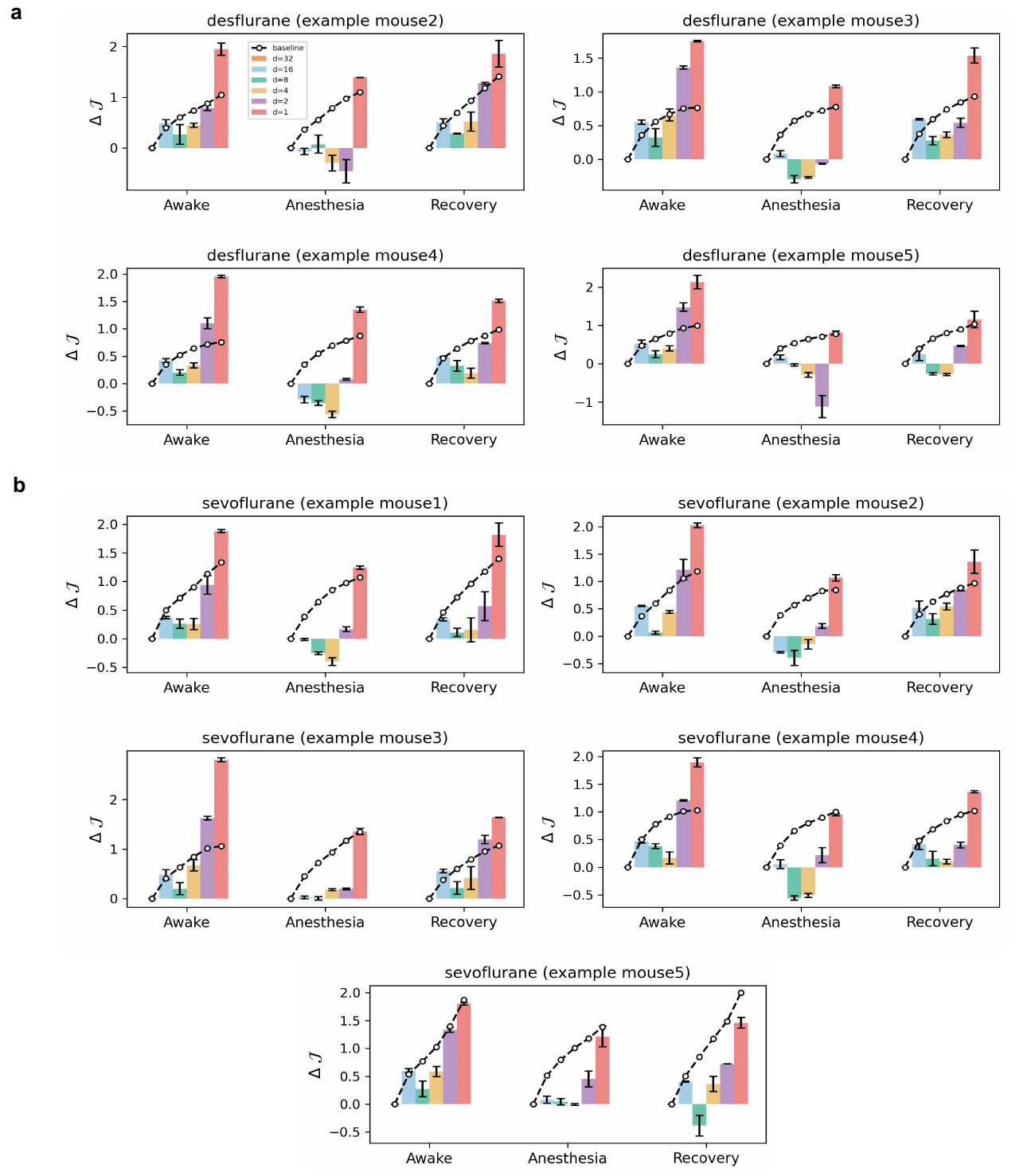}
    \caption{\textbf{The comparison of the degree
of causal emergence ($\Delta \mathcal{J}_d$) at different dimensions in three different stages}. Here, “baseline” represents the results when the mice exhibit Brownian motion dynamics. \textbf{a} The results in desflurane from mouse2 to mouse5. \textbf{b} The results in sevoflurane from mouse1 to mouse5. The results are the average of three repeated experiments for the original data and ten repetitions for the shuffled data, and the error bar represents the standard deviation.}
    \label{fig:si_ce_different_state}
\end{figure}

Fig.~\ref{fig:si_ei_different_state} confirms that $\mathcal{J}_1$ at $d$=1 reliably quantifies causal effects: highest for the original data, lower for shuffled data, and lowest for the baseline. Fig.~\ref{fig:si_ce_different_state} examines the $\Delta \mathcal{J}_d$ across different scales. In all three stages, the degree of causal emergence ($\Delta \mathcal{J}_1$) peaks at $d$=1, while a baseline model simulating
Brownian motion yields a much lower $\Delta \mathcal{J}_1$ compared to the real data, particularly during the awake stage.  These findings hold across different mice and anesthetics.

\section{The comparison of causal contribution distribution at different dimensions under two different anesthetics}\label{sec:si_causal_dis}

\begin{figure}[!ht]
    \centering
    \includegraphics[width=0.8\textwidth]{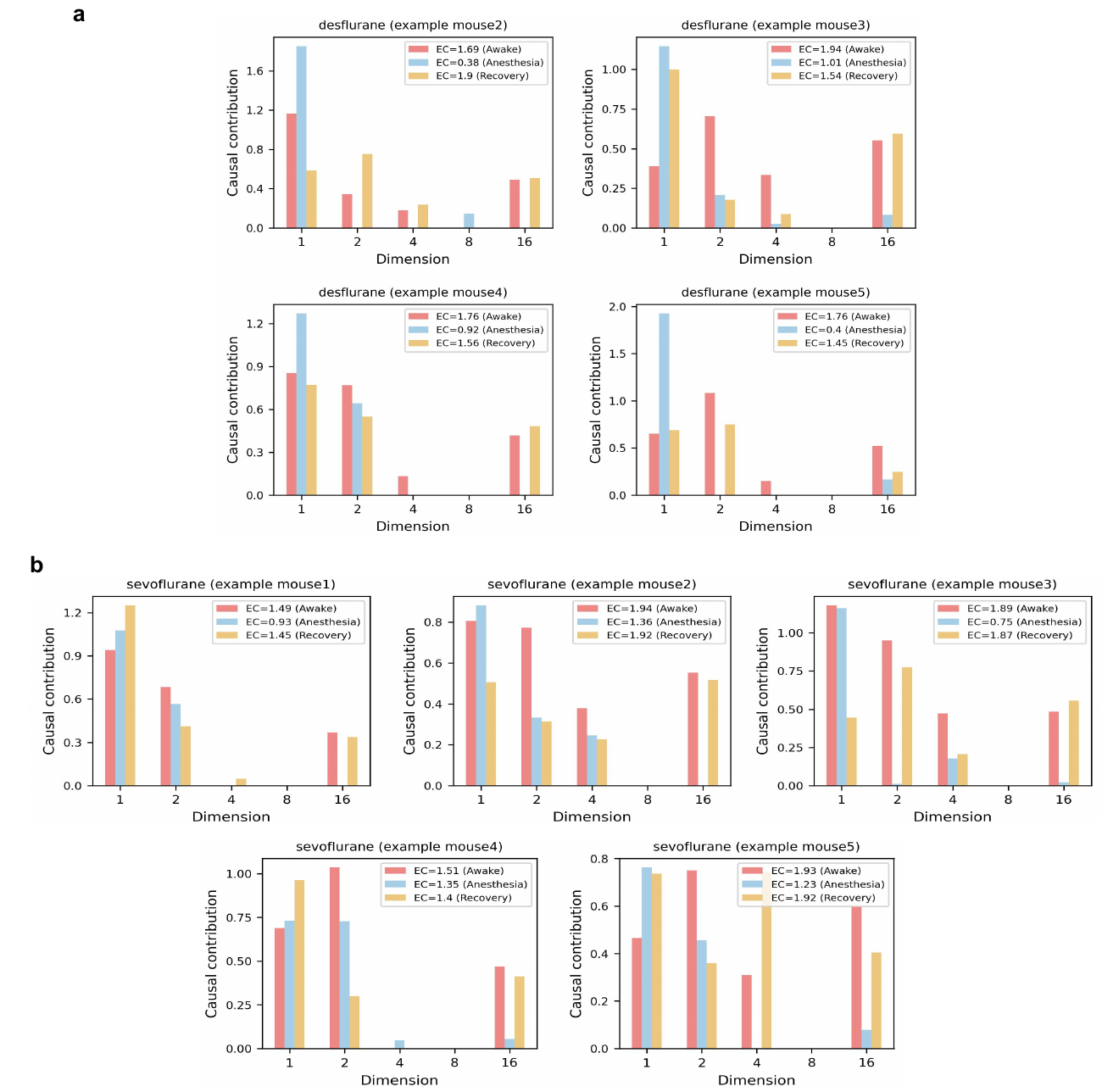}
    \caption{\textbf{The distribution of causal contributions at different dimensions in three stages.} \textbf{a} The results in desflurane from mouse2 to mouse5. \textbf{b} The results in sevoflurane from mouse1 to mouse5. Emergent complexity (EC) is calculated based on causal contribution ($q_k$) distribution~where $q_k=\max(\mathcal{J}_{2^k}-\mathcal{J}_{2^{k+1}},0)$ for $k \in \left\{0,1,2,3,4\right\}$, using Shannon entropy~($H=-\sum_{k}m_k\log_2 {m_k}, m_k= \frac{q_k}{\sum_{k}q_k}$).}
    \label{fig:si_cauasl_distribution}
\end{figure}

To more meticulously and comprehensively capture the marginal gain of causal power as the scale increases, we computed the causal contribution distribution for each dimension and, based on this, derived the "Emergent Complexity (EC)" as a statistical measure to quantify differences across the three conscious stages~\cite{hoel2025causal}. Fig.~\ref{fig:si_cauasl_distribution}a displays the causal contribution distribution under desflurane for different mice: the awake and recovery periods show a more uniform distribution compared to the anesthetized period, thus exhibiting higher emergent complexity. T-tests indicate significant differences between stages (awake vs. anesthetized: $t=4.81, p=0.009$; awake vs. recovery: $t=1.46, p=0.22$; recovery vs. anesthetized: $t=3.42, p=0.027$). Furthermore, Fig.~\ref{fig:si_cauasl_distribution}b shows results under sevoflurane for different mice, consistent with the conclusions from Fig.~\ref{fig:si_cauasl_distribution}a (awake vs. anesthetized: $t=2.52, p=0.06$; awake vs. recovery: $t=0.52, p=0.62$; recovery vs. anesthetized: $t=2.33, p=0.08$).

\section{Analysis of Emergent Dynamics}\label{sec:si_dynamics_analyse}

\begin{figure}[!ht]
    \centering
    \includegraphics[width=1\textwidth]{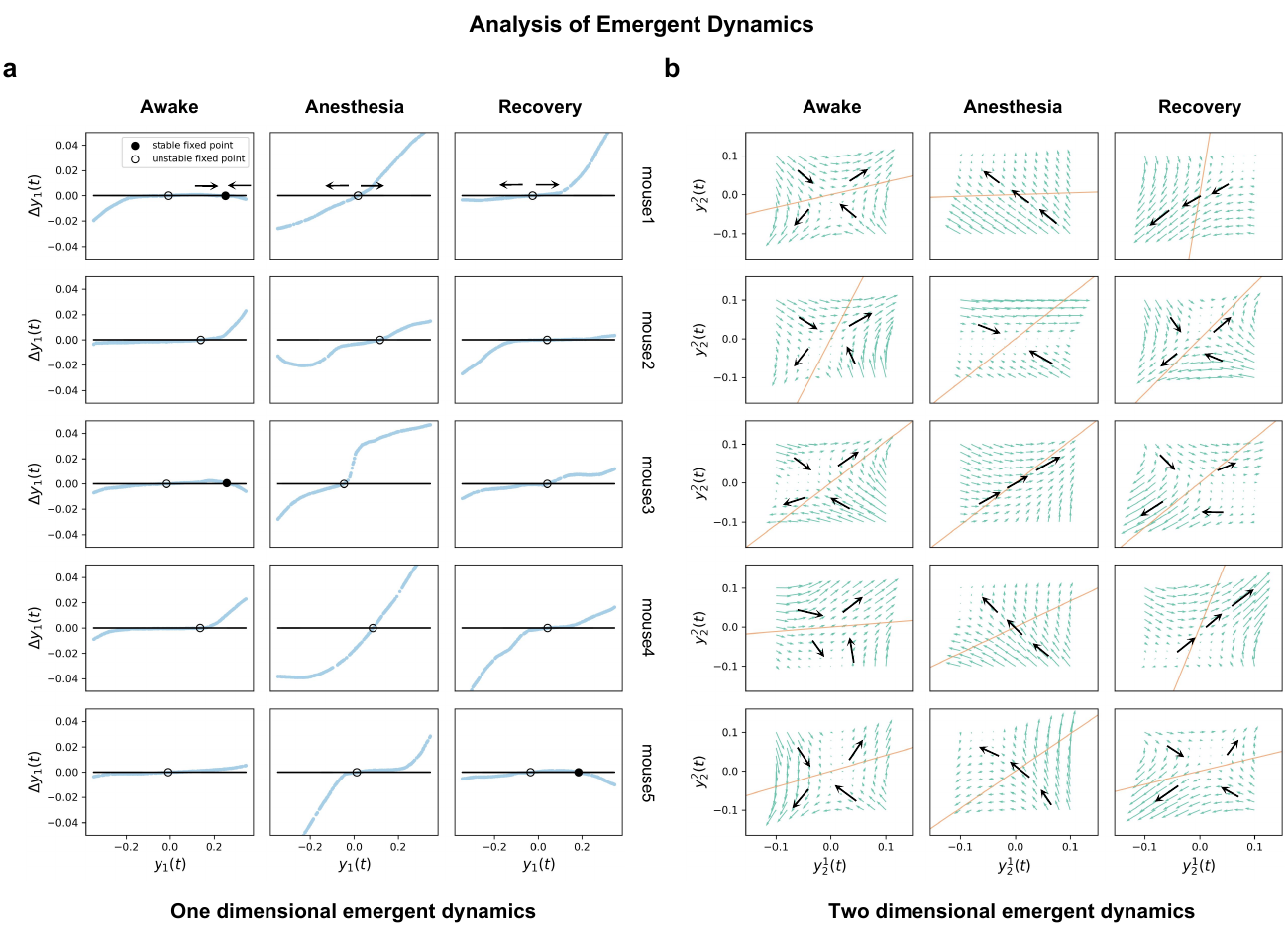}
    \caption{\textbf{The analysis of macro-dynamics of one-dimensional ($\boldsymbol{y}_1(t)$) and two-dimensional ($\boldsymbol{y}_2(t)$) latent variables for five mice across three different stages, specifically examining the dynamical neural networks $f_d$ for $d=1,2$ (sevoflurane).}  \textbf{a} Phase diagram of the one-dimensional macro-dynamics, illustrating the relationship between the current state $\boldsymbol{y}_1(t)$ and the predicted change $\Delta \boldsymbol{y}_1(t)$ ($\boldsymbol{\hat{y}}_1(t+1)-\boldsymbol{y}_1(t)$) at the next time step. Filled points represent stable fixed points, and hollow points represent unstable fixed points. The horizontal black line corresponds to $\Delta \boldsymbol{y}_1(t)=0$. The arrow in the panel indicates the direction of movement when the variable is in this state. \textbf{b} Vector field representation of the two-dimensional dynamics (see~\cite{yang2025finding} for details). Each green arrow indicates the direction and magnitude of the dynamical derivative ($dy^1_2/dt,dy^2_2/dt$) at the corresponding coordinate points. Black arrows illustrate the overall motion tendency of the vector field, and the orange line indicates the projection direction that approximately reduces the two-dimensional vector field to the one-dimensional dynamics.}
    \label{fig:si_dyna_analyse_one_two_dim_sev}
\end{figure}

The figure below illustrates the relationship between the learned one-dimensional and two-dimensional dynamics under sevoflurane. Fig.~\ref{fig:si_dyna_analyse_one_two_dim_sev}a illustrates the learned one-dimensional macro-dynamics. By comparing the relationship between the macro-variable $\boldsymbol{y}_1(t)$ and its change $\Delta \boldsymbol{y}_1(t)$, a consistent pattern was observed across different mice. In the awake stage, the dynamics exhibit a plateau with slow changes. The system can stabilize at this point or switch between states under noise or external perturbations, consistent with the ability of awake mice to flexibly respond to external stimuli and internal signals~\cite{mcginley2015waking,vyazovskiy2014dynamics,hancock2025metastability}. Regions beyond the plateau showing upward or downward bending correspond to unstable states. Under anesthesia, only unstable fixed points are observed, with a significantly larger numerical range, indicating that the conscious variable tends to diverge monotonically—a hallmark of loss of consciousness. The recovery stage exhibits intermediate characteristics: a plateau with slow changes emerges, but an unstable fixed point remains, suggesting that the system retains some destabilization, though substantially reduced compared to the anesthetized stage.

Fig.~\ref{fig:si_dyna_analyse_one_two_dim_sev}b visualizes the learned dynamics of two-dimensional macro-variables through a vector field representation. In the awake stage, the mice consistently exhibit saddle points, which is a characteristic of consciousness \cite{hancock2025metastability,finkelstein2021attractor,tognoli2014metastable,rabinovich2011robust}. Under anesthesia, the dynamics show unstable unidirectional motion, indicating a destabilized state. The recovery lies between these two stages: saddle points occasionally emerge, but unstable unidirectional motion also occurs. By projecting the two-dimensional vector field along the orange line (projection direction), it can be approximately aligned with the one-dimensional dynamics shown in Fig.~\ref{fig:si_dyna_analyse_one_two_dim_sev}a.

\begin{figure}[!ht]
    \centering
    \includegraphics[width=1\textwidth]{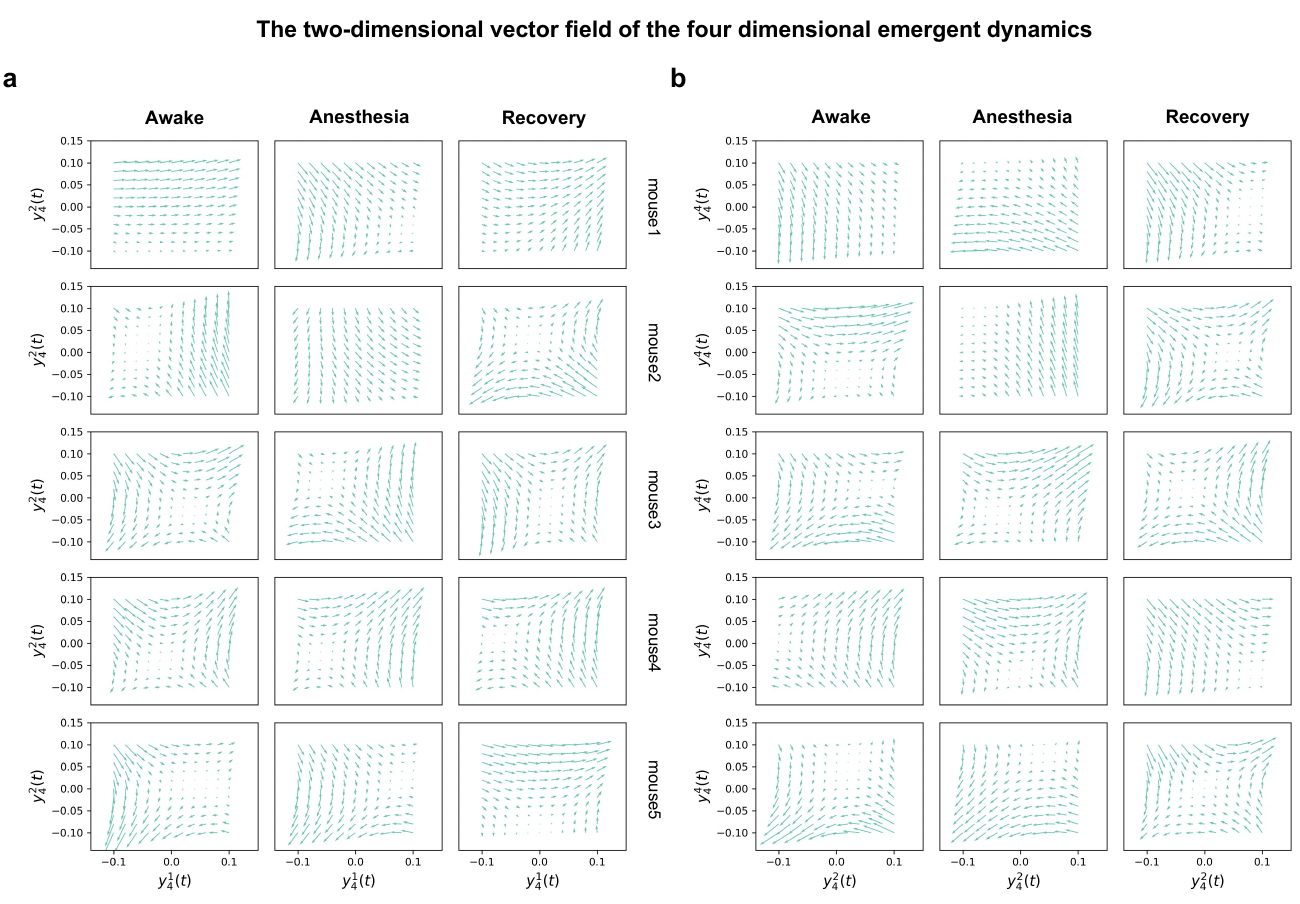}
    \caption{\textbf{The analysis of emergent dynamics in different stages based on dimension four (desflurane).} \textbf{a} Each subplot represents the two-dimensional vector field of dynamics between the first and second dimensions of $d$=4. \textbf{b} Each subplot represents the two-dimensional vector field of dynamics between the second and fourth dimensions of $d$=4.}
    \label{fig:si_dyna_analyse_four_dim_des}
\end{figure}

Since visualizing dynamics beyond two dimensions is challenging, we simplified the representation of the four-dimensional dynamics by selecting two dimensions for vector field visualization. Specifically, we chose the first and second dimensions, as well as the second and fourth dimensions under desflurane, as shown in Fig.~\ref{fig:si_dyna_analyse_four_dim_des}a and ~\ref{fig:si_dyna_analyse_four_dim_des}b. The results indicate that both plots consistently exhibit saddle-point dynamics across all three stages, failing to correspond to actual conscious states. Thus, this scale can not reveal unique dynamical mechanisms distinguishing awake from anesthetized conditions. Similar patterns were observed for any other pairwise combination of the four dimensions, and no additional visualizations are provided.

The figure below displays two-dimensional vector field visualizations of the four-dimensional dynamics under sevoflurane. The settings are consistent with Fig.~\ref{fig:si_dyna_analyse_four_dim_des}, and the conclusions remain unchanged: the dynamics at this scale fail to reveal unique mechanisms distinguishing awake from anesthetized stages.

\begin{figure}[!ht]
    \centering
    \includegraphics[width=0.9\textwidth]{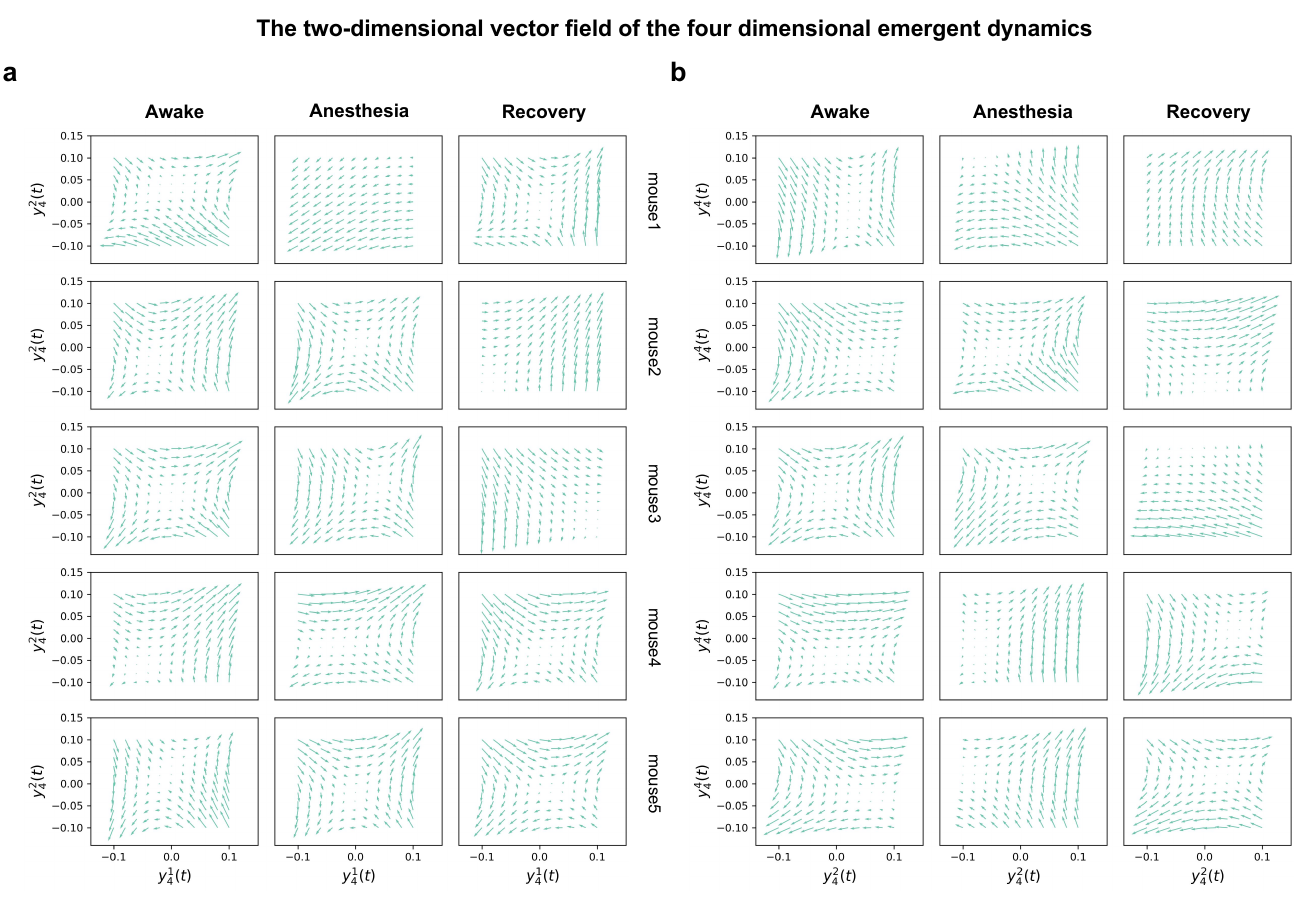}
    \caption{\textbf{The analysis of emergent dynamics in different stages based on dimension four (sevoflurane).} \textbf{a} Each subplot represents the two-dimensional vector field of dynamics between the first and second dimensions of $d$=4. \textbf{b} Each subplot represents the two-dimensional vector field of dynamics between the second and fourth dimensions of $d$=4.}
    \label{fig:si_dyna_analyse_four_dim_sev}
\end{figure}

\section{Hypothetical neighborhood-based grouping strategy}\label{sec:si_hypothetical_grouping_strategy}

\begin{figure}[!ht]
    \centering
    \includegraphics[width=0.8\textwidth]{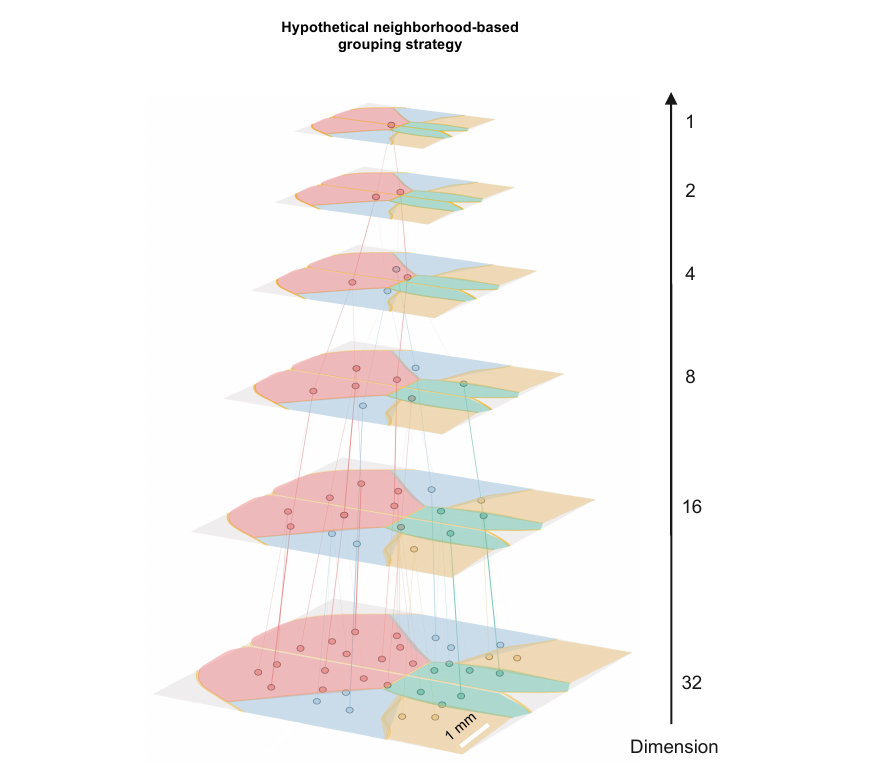}
    \caption{\textbf{The hypothetical neighborhood-based information aggregation strategy.} Nodes at the bottom layer ($d=32$) represent ``neuronal functional ensembles''. The position of each node above 16 dimensions corresponds to the centroid of the cluster formed by the nodes it covers in the layer below, where clustering at each layer is performed using K-means based on spatial coordinates. The weights of downward edges from each variable are randomly sampled from a uniform distribution ($U(0,1$)).}
    \label{fig:si_hypothetical_grouping_strategy}
\end{figure}

\section{The analysis of coverage across different dimensions}\label{sec:si_coverage}

Fig.~\ref{fig:si_attribution_analyse} shows the change in the coverage of information integration in the mouse brain across dimensions (from $d$=16 to $d$=1). The real brain exhibits a large coverage even at the initial period of micro-scale integration, which is significantly higher than that of the hypothetical neighborhood way, indicating distinct integration mechanisms between these two methods. The coverage increases as the dimension decreases, suggesting that lower-dimensional causal variables require the integration of more microscopic information to represent complex macro-states. The coverage during the awake stage is significantly higher than during anesthesia/recovery, implying that conscious activity relies on broader neuronal information integration. These findings remain consistent across different mice and anesthetics.

\begin{figure}[!ht]
    \centering
    \includegraphics[width=0.9\textwidth]{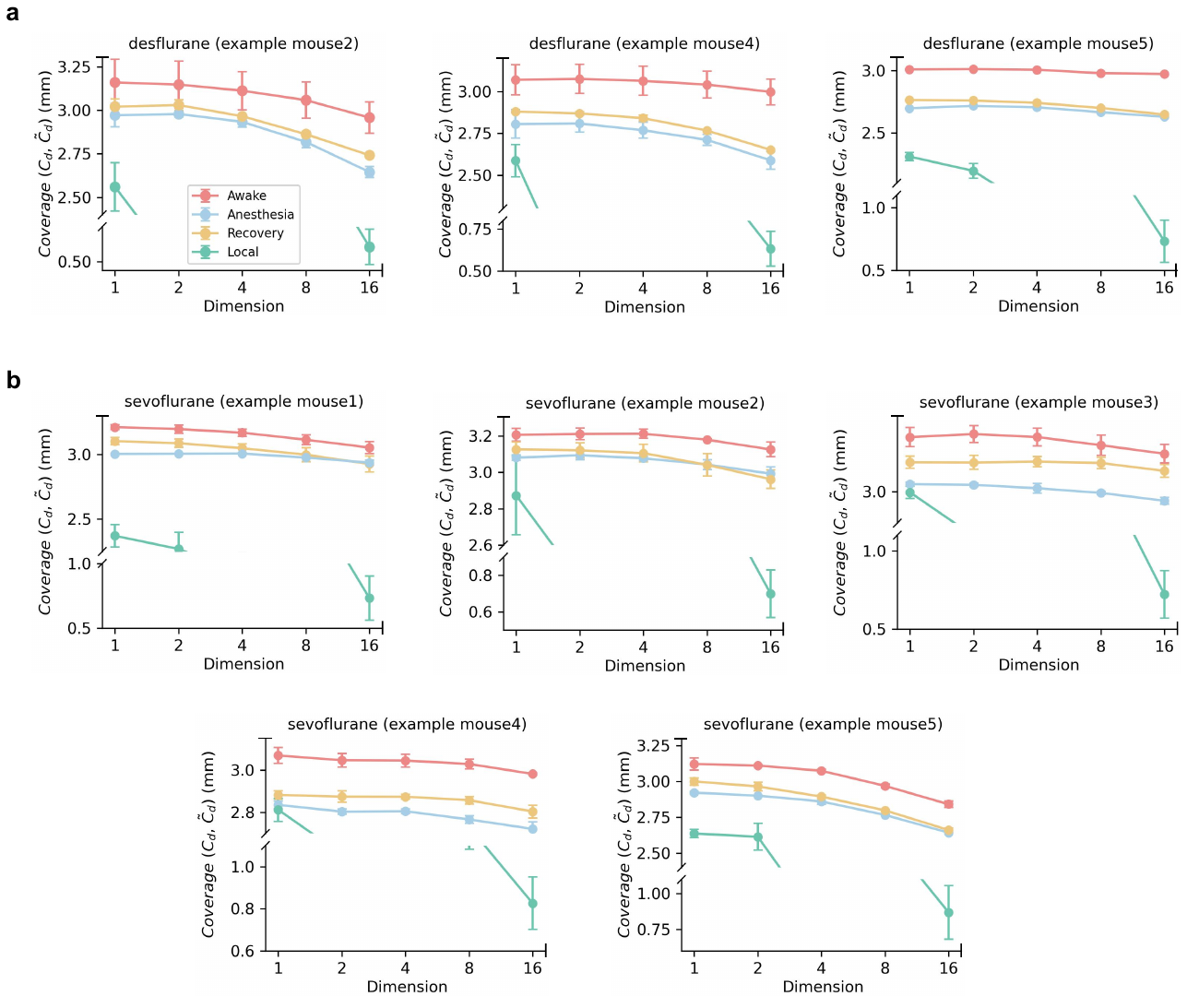}
    \caption{\textbf{The approximate coverage of significant attribution areas in the bottom level corresponding to the high-level variables under both approaches, including the results from the NIS+ model and the hypothetical neighborhood-based way.} \textbf{a} The results in desflurane from mouse2 to mouse5. \textbf{b} The results in sevoflurane from mouse1 to mouse5. See Sec.~\ref{cal_coverage} for detailed calculations of coverage. The results are the average of three repeated experiments, and the error bar represents the standard deviation.}
    \label{fig:si_coverage}
\end{figure}

\section{Attribution to brain regions}\label{sec:si_attri_analyse}

\begin{figure}[!ht]
    \centering
    \includegraphics[width=1\textwidth]{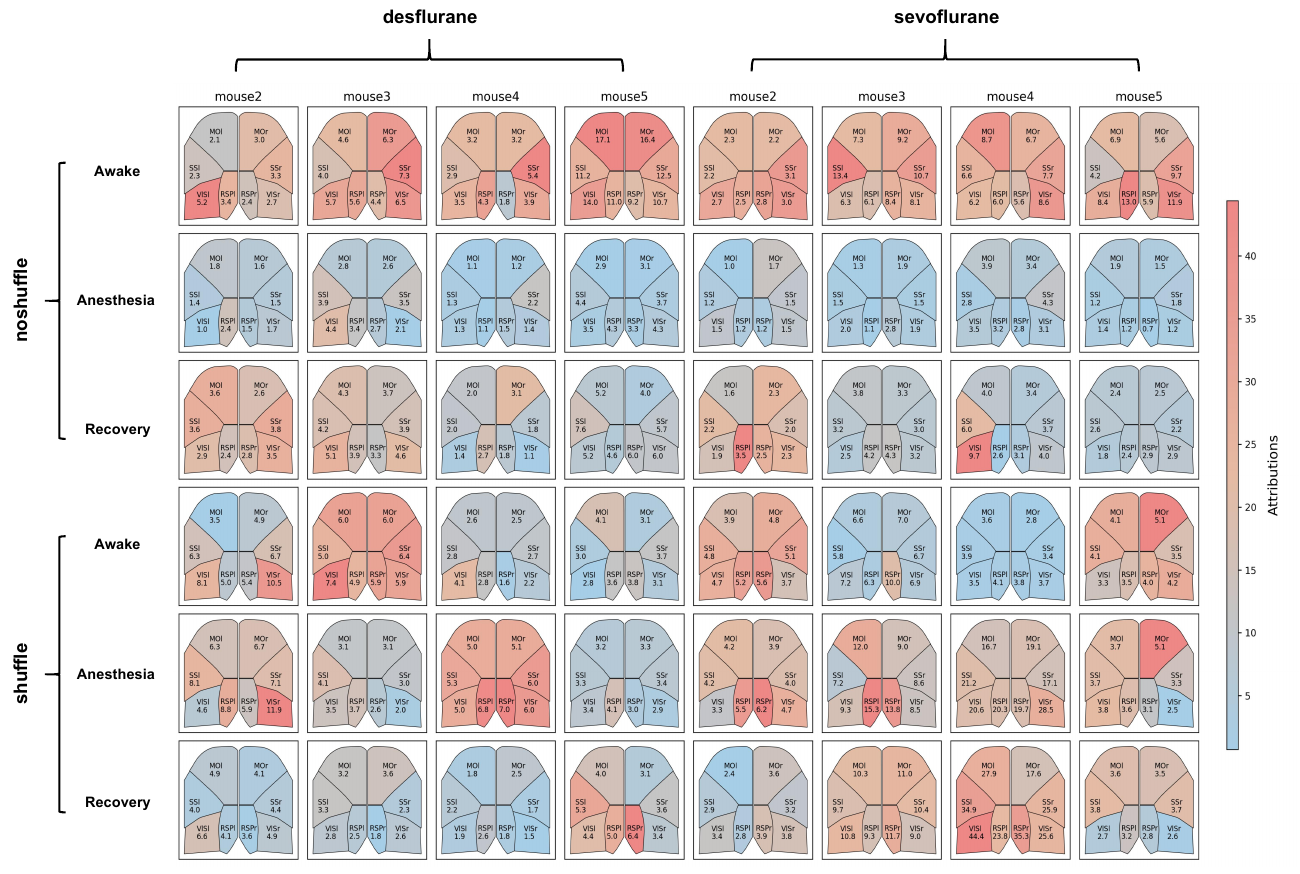}
    \caption{\textbf{The comparison of brain region attribution calculated from the top-level one-dimensional variable to the bottom dimension between shuffled and non-shuffled cases in three different stages~(devflurane vs sevoflurane, from mouse2 to mouse5).} The attribution value for each subplot is obtained by summing the attribution values of all neurons within the same brain region, denoted as $A_b$, where $b$ represents the corresponding brain region, and the attribution value of a neuron equals that of its corresponding ensemble. The attribution values reported in the panel are the mean of three repeated experiments for the original data and ten repetitions for the shuffled data, amplified by a factor of 1000.}
    \label{fig:si_attribution_analyse}
\end{figure}

Fig.~\ref{fig:si_attribution_analyse} presents the brain region attribution results across different mice under two anesthetics, with both shuffled and non-shuffled data, and throughout three experimental stages (awake, anesthetized, recovery). By identifying micro-scale neuronal functional ensembles that significantly influence the macroscopic conscious variable, the attribution analysis reveals distinct patterns of regional contributions across conscious states---most regions exhibit $A_b$(awake) $>$ $A_b$(recovery) $>$ $A_b$(anesthetized)---and further identifies key brain areas associated with consciousness (such as MOr, SSl, and VIS).

\section{Model Parameter}
\label{model_parameter}
In this section, we outline the hyperparameters employed during model training, allowing interested readers to replicate our experimental results. In general, our model shows limited sensitivity to hyperparameters.

\begin{itemize}
    \item Hyperparameter Settings for Model
    \begin{itemize}
        \item Epochs: Stage one: 10000; Stage two: 50000
        \item Activation Function: LeakyReLU Function
        \item Hidden Layers for Forward Dynamic and Inverse Dynamic Learners: 5
        \item Hidden Units: 100 
        \item Layers of  RealNVP~\cite{dinh2016density}~(A type of Invertible neural network) modules in Encoder~(Decoder): 6 
        \item Batch Size: 256
        \item L: 1 
        \item Learning Rate: fixed to 0.0001
        \item Optimizer: Adam
        \item Weight for Loss Function ($\lambda$): 1 (To treat both error terms equally) 
    \end{itemize}
\end{itemize}

\end{appendices}

\end{document}